\documentclass[aps,twocolumn]{revtex4}

\RequirePackage[utf8]{inputenc}
\RequirePackage[T2A]{fontenc}

\usepackage{amssymb}
\usepackage{amsmath}
\usepackage{bm}
\usepackage[dvips]{epsfig}
\usepackage{graphicx}
\usepackage{textcomp}

\begin{document}

\title{Analytical approximations of dispersion laws and 
ultra-complex conductivity diagrams} 

\author{A.Ya. Maltsev}

\affiliation{
\begin{center}
{\it Steklov Mathematical Institute of Russian Academy 
of Sciences}
\end{center}
\centerline{\it 8 Gubkina St., Moscow 119333, Russia}
\begin{center}
{\it L.D. Landau Institute for Theoretical Physics
of Russian Academy of Sciences}
\end{center}
\centerline{\it 142432 Chernogolovka, pr. Ak. Semenova 1A}
}

\begin{abstract}
 We study the probability of the emergence of ultra-complex 
conductivity diagrams in conductors that satisfy the tight-binding 
approximation and have the simple or body-centered cubic lattice. 
The presence of ultra-complex conductivity diagrams allows us 
to observe a number of highly nontrivial effects in strong 
magnetic fields, however, the probability of their emergence
in a given substance is quite low. In the case of the simple 
or body-centered cubic lattice, the leading tight-binding 
approximation does not allow us to estimate this probability 
due to the peculiarities of the spectra in this situation. 
To estimate this probability, we use higher-order corrections 
to the leading approximation, which yield more accurate 
analytical expressions for the electron spectra.
\end{abstract}

\maketitle

\section{Introduction}

 In this paper, we consider the dispersion relations arising 
in crystals of cubic symmetry in the tight-binding approximation. 
As is well known, the electron states in crystals are determined 
by the energy band number and the value of the quasimomentum 
$\, {\bf p} \in \mathbb{R}^{3} \, $, determined up to the vectors 
of the reciprocal lattice $\, L^{*} \, $. The basis 
$\, \left( {\bf a}_{1}, {\bf a}_{2}, {\bf a}_{3} \right) \, $ 
of the lattice $\, L^{*} \, $ is given by the relations
\begin{multline*}
{\bf a}_{1} \,\,\, = \,\,\, 2 \pi \hbar \,\,
{{\bf l}_{2} \, \times \, {\bf l}_{3} \over
({\bf l}_{1}, \, {\bf l}_{2}, \, {\bf l}_{3} )} \,\,\, , \quad \quad
{\bf a}_{2} \,\,\, = \,\,\, 2 \pi \hbar \,\,
{{\bf l}_{3} \, \times \, {\bf l}_{1} \over
({\bf l}_{1}, \, {\bf l}_{2}, \, {\bf l}_{3} )} \,\,\, ,  \\
{\bf a}_{3} \,\,\, = \,\,\, 2 \pi \hbar \,\,
{{\bf l}_{1} \, \times \, {\bf l}_{2} \over
({\bf l}_{1}, \, {\bf l}_{2}, \, {\bf l}_{3} )} \,\,\, , 
\quad \quad \quad \quad
\end{multline*}
where $\, \left( {\bf l}_{1}, {\bf l}_{2}, {\bf l}_{3} \right) \, $ 
represent the basis of the crystallographic lattice $\, L \, $.

 Thus, all the vectors
\begin{multline*}
{\bf p} \,\,\, \equiv \,\,\, {\bf p} \,\, + \,\, 
n_{1} \, {\bf a}_{1} \,\, + \,\, n_{2} \, {\bf a}_{2} \,\, + \,\,
n_{3} \, {\bf a}_{3} \,\,\, ,  \\
n_{1} , \, n_{2} , \, n_{3} \,\, \in \,\, \mathbb{Z} 
\end{multline*}
define the same electronic state for a given energy band. 
As a consequence, the values of the electron quasi-momentum 
can be considered as points on a three-dimensional torus
$${\bf p} \,\,\, \in \,\,\, 
\mathbb{T}^{3} \,\,\, = \,\,\, \mathbb{R}^{3} \Big/ L^{*} 
\,\,\, , $$
obtained by factorizing the space $\, \mathbb{R}^{3} \, $
by the reciprocal lattice vectors.

 In practice, it is actually convenient to consider both 
representations $\, {\bf p} \in \mathbb{T}^{3} \, $ and 
$\, {\bf p} \in \mathbb{R}^{3} \, $.

 The dependence of the electron energy $\, \epsilon ({\bf p}) \, $
on its quasimomentum determines the dispersion relation 
(dispersion law) for a given energy band in a crystal. 
Depending on the representation, $\, \epsilon ({\bf p}) \, $ 
can be considered either a function on the torus 
$\, \mathbb{T}^{3} = \mathbb{R}^{3} \big/ L^{*} \, $, 
or a 3-periodic function in the full space $\, \mathbb{R}^{3} \, $.

\begin{figure*}[t]
\begin{center}
\includegraphics[width=\linewidth]{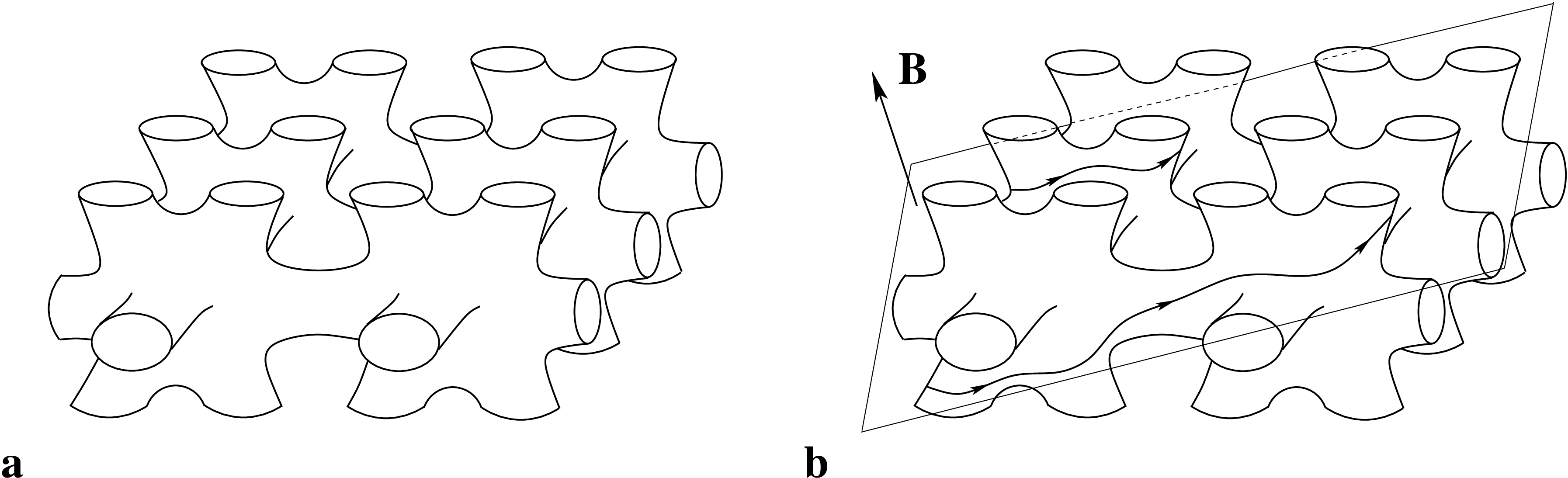}
\end{center}
\caption{(a) The Fermi surface of a complex shape 
in $\, {\bf p}$ - space. (b) Trajectories of 
system (\ref{MFSyst}) on the Fermi surface.}
\label{Fig1}
\end{figure*}

 As is well known, the greatest interest in the study of conductors 
is in the conduction bands, i.e., energy bands only partially filled 
with electrons. It is also well known that every conductor has the 
Fermi level (Fermi energy) $\, \epsilon_{F} \, $, separating 
occupied electron states from unoccupied ones. Each conduction 
band has its own Fermi surface
$$S_{F} \, : \quad \epsilon ({\bf p}) \, = \, 
\epsilon_{F}  \,\,\, , $$
which can also be viewed either as a smooth compact surface in 
the torus $\, \mathbb{T}^{3} \, $, or as a 3-periodic surface in 
the full $\, {\bf p}$ -space (Fig. \ref{Fig1}, a). Most electronic 
phenomena in conductors are determined by the behavior of electrons 
near the Fermi surface.

\vspace{1mm}

 This work is concerned with the description of electron 
transport phenomena in conductors in the presence of strong 
magnetic fields. The presence of an external magnetic field 
$\, {\bf B} \, $ causes the evolution of electron states 
in $\, {\bf p}$ - space according to the dynamic system
\begin{equation}
\label{MFSyst}
{\dot {\bf p}} \,\,\,\, = \,\,\,\, {e \over c} \,\,
\left[ {\bf v}_{\rm gr} ({\bf p}) \times {\bf B} \right]
\,\,\,\, \equiv \,\,\,\, 
{e \over c} \,\, \left[ \nabla \epsilon ({\bf p})
\times {\bf B} \right] 
\end{equation}
(see for example \cite{Kittel, Ziman, AshcroftMermin, Abrikosov}).

 Geometrically, the trajectories of system (\ref{MFSyst}) 
are defined by the intersections of planes orthogonal to 
$\, {\bf B} \, $, and constant energy surfaces 
$\, \epsilon ({\bf p}) = {\rm const} \, $ (Fig. \ref{Fig1}, b).

 The condition for strong magnetic fields can be formally 
written as $\, \omega_{B} \tau \gg 1 \, $, where 
$\, \omega_{B} = eB / m^{*} c \, $ is the cyclotron frequency, 
and $\, \tau \, $ is the electron mean free time ($m^{*}$ plays 
the role of the effective electron mass in the crystal).

 More precisely, the quantity $\, B \, $ is sufficiently large 
if the electron manages to travel a sufficiently large distance 
(much larger than the size of the Brillouin zone) along the 
trajectories of system (\ref{MFSyst}) between two scattering 
events. It should be noted immediately that such a situation 
arises only for sufficiently pure single-crystal samples at 
low temperatures ($T \leq 1 {\rm K}$) in sufficiently strong 
magnetic fields ($B \geq 1 {\rm Tl}$)

 As can be seen, for sufficiently complex Fermi surfaces, 
the shape of the trajectories of (\ref{MFSyst}) can also be 
very complex (Fig. \ref{Fig1}, b). In particular, the 
trajectories of (\ref{MFSyst}) can be both closed and open 
(unclosed) in the $\, {\bf p}$ - space (and in the torus 
$\, \mathbb{T}^{3} = \mathbb{R}^{3} \big/ L^{*} $). As was 
shown in \cite{lifazkag,lifpes1,lifpes2,etm}, the behavior 
of electron transport phenomena in the limit 
$\, \omega_{B} \tau \rightarrow \infty \, $ strongly depends 
on the shape of the trajectories of (\ref{MFSyst}) and, 
in particular, on the presence or absence of open trajectories 
on the Fermi surface.

 The shape of the trajectories of (\ref{MFSyst}) on a given 
Fermi surface depends significantly on the direction of 
$\, {\bf B} \, $. Here we will be interested in the behavior 
of the electrical conductivity tensor 
$\, \sigma^{kl} ({\bf B}) \, $ in the limit 
$\, B \rightarrow \infty \, $. In view of the above, 
the ``angular diagrams'' marking the emergence of open 
trajectories on the Fermi surface (as well as their type) 
for different directions 
$${\bf n} \,\,\, = \,\,\, {\bf B} / B \,\,\, \in \,\,\, 
\mathbb{S}^{2} $$
will be very important for us.

 The problem of describing all possible types of trajectories 
of system (\ref{MFSyst}) (for an arbitrary dispersion relation 
$\, \epsilon ({\bf p})$) was set by S.P. Novikov in his paper 
\cite{MultValAnMorseTheory}. This problem, like its numerous 
applications, was actively studied in his topological school 
(see, e.g., \cite{zorich1,dynn1992,Tsarev,dynn1,zorich2,DynnBuDA,dynn2,
dynn3,PismaZhETF,ZhETF1997,UFN,BullBrazMathSoc,StatPhys,DeLeoPhysLettA,
DeLeoPhysB,TrMIAN,DeLeo2017,SecBound,UltraCompl,DynMalNovUMN}). 
In particular, a complete classification of all possible types of 
open trajectories of (\ref{MFSyst}) (as well as the corresponding 
behavioral regimes of the tensor $\, \sigma^{kl} ({\bf B}) \, $ for 
$\, B \rightarrow \infty $) has now been obtained. Here we will use 
this classification in describing the above-mentioned angular diagrams. 
We will also call such diagrams angular diagrams of conductivity 
(of a metal) in strong magnetic fields.

\begin{figure*}[t]
\begin{center}
\includegraphics[width=\linewidth]{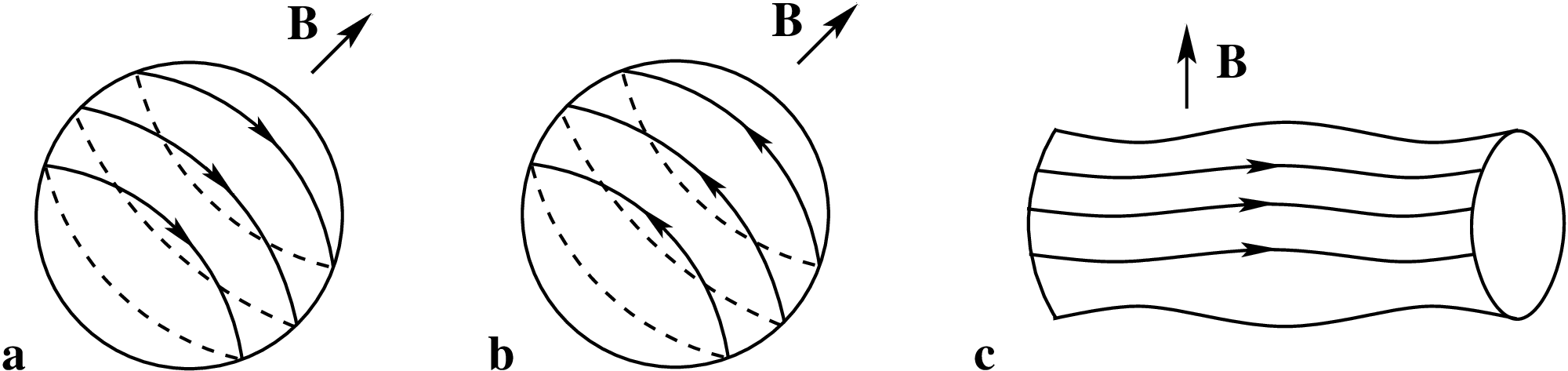}
\end{center}
\caption{(a,b) Simple Fermi surfaces of electron and hole 
type carrying closed trajectories of system (\ref{MFSyst}). 
(c) Periodic trajectories of system (\ref{MFSyst}).}
\label{Fig2}
\end{figure*}

 For simple Fermi surfaces (Fig. \ref{Fig2}, a, b), the 
angular conductivity diagrams are also very simple. Namely, 
each direction of $\, {\bf B} \, $ corresponds to the presence 
of only closed trajectories (of a fairly simple shape) on the 
Fermi surface. The behavior of the conductivity tensor in 
strong magnetic fields is described by the formula
(\cite{lifazkag}):
\begin{equation}
\label{ClosedTr}
\sigma^{kl}_{\rm closed} \,\,\,\, \simeq \,\,\,\,
{n e^{2} \tau \over m^{*}} \, \left(
\begin{array}{ccc}
( \omega_{B} \tau )^{-2}  &  ( \omega_{B} \tau )^{-1}  &
( \omega_{B} \tau )^{-1}  \cr
( \omega_{B} \tau )^{-1}  &  ( \omega_{B} \tau )^{-2}  &
( \omega_{B} \tau )^{-1}  \cr
( \omega_{B} \tau )^{-1}  &  ( \omega_{B} \tau )^{-1}  &  *
\end{array}  \right) \,\,\, ,   
\end{equation}
($\omega_{B} \tau \rightarrow \infty $).

 Formula (\ref{ClosedTr}) describes the asymptotic behavior 
of the components $\, \sigma^{kl} ({\bf B}) \, $ in the limit 
$\, B \rightarrow \infty \, $; in particular, it is assumed 
that each matrix element in (\ref{ClosedTr}) has a dimensionless 
factor of order 1. The quantity $\, n \, $ here represents 
the concentration of charge carriers in the metal (and the 
sign $\, * \, $ denotes a dimensionless quantity of order 1). 
In formula (\ref{ClosedTr}), it is also assumed that the 
$\, z \, $ axis is directed along the $\, {\bf B} \, $ direction.

\vspace{1mm}

 An important remark should be made immediately, however. 
``Simple'' Fermi surfaces can be divided into two types, 
namely, electron-type Fermi surfaces (Fig. \ref{Fig2}, a) 
and hole-type Fermi surfaces (Fig. \ref{Fig2}, b). 
Electron-type Fermi surfaces enclose regions
$\, \epsilon ({\bf p}) < \epsilon_{F} \, $, while hole-type 
surfaces enclose regions $\, \epsilon ({\bf p}) > \epsilon_{F} \, $, 
which is directly reflected in the behavior of the Hall 
conductivity $\, \sigma^{xy} ({\bf B}) \, $. Namely, 
the contribution of the electron-type Fermi surface to 
the Hall conductivity is given in the leading order by the 
formula (\cite{Kittel, Ziman, AshcroftMermin, Abrikosov, etm})
\begin{equation}
\label{ElType}
\sigma^{xy} \,\,\, = \,\,\, - \sigma^{yx} \,\,\, = \,\,\,  
{e c \over B} \,\, {2 V \over (2 \pi \hbar)^{3}} \,\,\, ,  
\end{equation}
while the contribution from the hole-type Fermi surface 
is given by the formula
\begin{equation}
\label{HoleType}
\sigma^{xy} \,\,\, = \,\,\, - \sigma^{yx} \,\,\, = \,\,\, - \,\, 
{e c \over B} \,\, {2 V \over (2 \pi \hbar)^{3}} \,\,\, ,  
\end{equation}

 In formulas (\ref{ElType}) - (\ref{HoleType}), the 
quantity $\, V \, $ represents the volume bounded by 
the Fermi surface in $\, {\bf p}$ - space. Note that 
the quantities (\ref{ElType}) - (\ref{HoleType}) do not 
depend on the direction of $\, {\bf B} \, $, i.e., they 
are constant over the entire angular diagram for a given 
value of $\, B \, $ (for ``simple'' Fermi surfaces).

 It can also be noted that, for the relation 
$\, \epsilon ({\bf p}) \, $ with the range of values
$$\epsilon_{\min} \,\,\, \leq \,\,\, \epsilon ({\bf p})
\,\,\, \leq \,\,\, \epsilon_{\max} \,\,\, , $$
electron-type Fermi surfaces arise at $\, \epsilon_{F} \, $ 
close to $\, \epsilon_{\min} \, $, and hole-type Fermi 
surfaces arise at $\, \epsilon_{F} \, $ close 
to $\, \epsilon_{\max} \, $.

\vspace{1mm}

 Periodic open trajectories (Fig. \ref{Fig2}, c) represent 
the simplest open trajectories of system (\ref{MFSyst}). 
Their contribution to the tensor $\, \sigma^{kl} ({\bf B}) \, $ 
in the limit $\, \omega_{B} \tau \rightarrow \infty \, $ 
is given by the asymptotic formula (\cite{lifazkag}):
\begin{equation}
\label{PeriodicTr}
\Delta \sigma^{kl}_{\rm periodic} \,\,\,\, \simeq \,\,\,\,
{n e^{2} \tau \over m^{*}} \, \left(
\begin{array}{ccc}
( \omega_{B} \tau )^{-2}  &  ( \omega_{B} \tau )^{-1}  &
( \omega_{B} \tau )^{-1}  \cr
( \omega_{B} \tau )^{-1}  &  *  &  *  \cr
( \omega_{B} \tau )^{-1}  &  *  &  *
\end{array}  \right)  
\end{equation}

 In formula (\ref{PeriodicTr}), as in formula (\ref{ClosedTr}), 
the $\, z \, $ axis is directed along the magnetic field 
$\, {\bf B} \, $. Moreover, in formula (\ref{PeriodicTr}) the 
direction of the $\, x \, $ axis coincides with the mean direction 
of the open trajectories in $\, {\bf p}$ - space. It is easy to 
see that the main difference between formula (\ref{PeriodicTr}) 
and formula (\ref{ClosedTr}) is the emergence of strong anisotropy 
of conductivity in the plane orthogonal to $\, {\bf B} \, $. 
Note also that when open trajectories appear on the Fermi surface, 
formulas (\ref{ElType}) - (\ref{HoleType}) cease to be satisfied.

 It can be seen that the periodic trajectories shown in 
Fig. \ref{Fig2}, c, are stable for small rotations of 
$\, {\bf B} \, $ in the plane orthogonal to their mean direction, 
and are destroyed for other rotations of $\, {\bf B} \, $. 
The emergence of such trajectories on the Fermi surface thus 
corresponds to one-dimensional segments on the angular 
diagram (i.e., on the sphere 
$\, \mathbb{S}^{2} \ni {\bf n} = {\bf B}/B$).

 It is easy to see that the emergence of open trajectories 
of system (\ref{MFSyst}) is possible only on sufficiently 
complex Fermi surfaces. In general, the complexity of a Fermi 
surface is characterized by its genus $\, g \, $, as well as 
its topological rank $\, {\rm Rank} \, S_{F} \, $.

\begin{figure*}[t]
\begin{center}
\includegraphics[width=\linewidth]{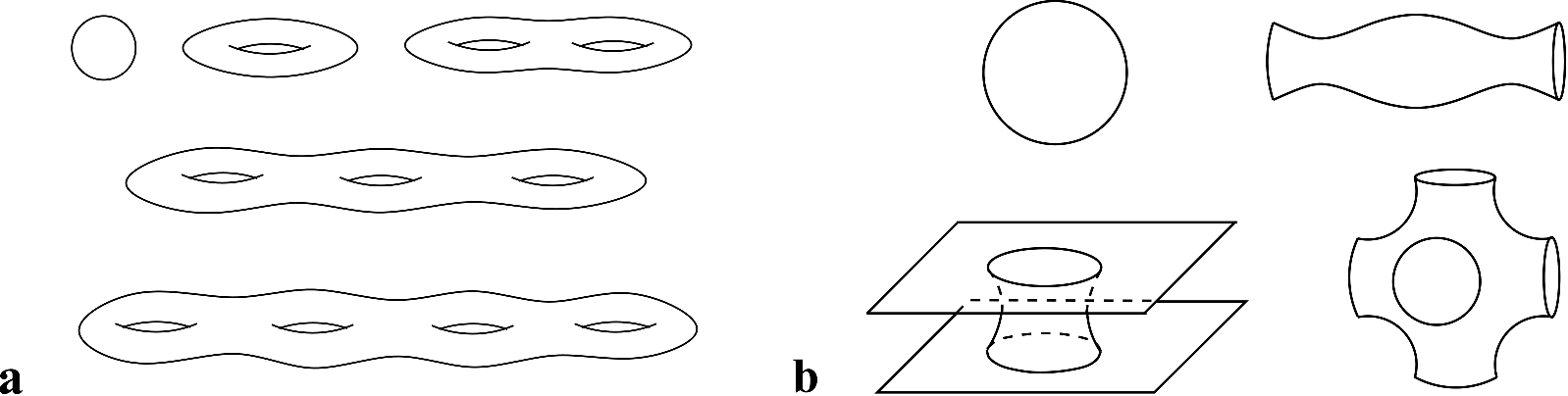}
\end{center}
\caption{(a) Abstract Fermi surfaces of genus 0, 1, 2, 3, and 4.
(b) Fermi surfaces of rank 0, 1, 2, and 3.}  
\label{Fig3}
\end{figure*}

 The genus of the Fermi surface is related to its 
representation as a compact (orientable) surface
$$S_{F} \,\,\, \subset \,\,\, \mathbb{T}^{3} \,\,\, = \,\,\,
\mathbb{R}^{3} \big/ L^{*} \,\,\, , $$
embedded in a three-dimensional torus. In this representation, 
the Fermi surface is homeomorphic to one of the canonical 
surfaces (Fig. \ref{Fig3}, a), determined by the value 
$\, g \, $. The genus of the Fermi surface can take the 
values $\, g \, = \, 0, 1, 2, 3, 4, \dots \, $.

 The rank of the Fermi surface is related to the features 
of its embedding in $\, \mathbb{T}^{3} \, $ and 
$\, \mathbb{R}^{3} \, $ and is equal to the number of independent 
directions in which it extends in $\, {\bf p}$ - space 
(Fig. \ref{Fig3}, b). The rank of the Fermi surface can take 
values 0, 1, 2, and 3.

 It is easy to see that the Fermi surfaces shown in 
Fig. \ref{Fig2}, a, b, have genus $\, g = 0 \, $ and 
$\, {\rm Rank} \, S_{F} = 0 \, $, while the Fermi surface 
shown in Fig. \ref{Fig2}, c, has genus $\, g = 1 \, $ 
and $\, {\rm Rank} \, S_{F} = 1 \, $. Here we will be 
interested in more complicated Fermi surfaces, having genus 
$\, g \geq 3 \, $ and $\, {\rm Rank} \, S_{F} \, = \, 3 \, $.

\vspace{1mm}

\begin{figure}[t]
\begin{center}
\includegraphics[width=\linewidth]{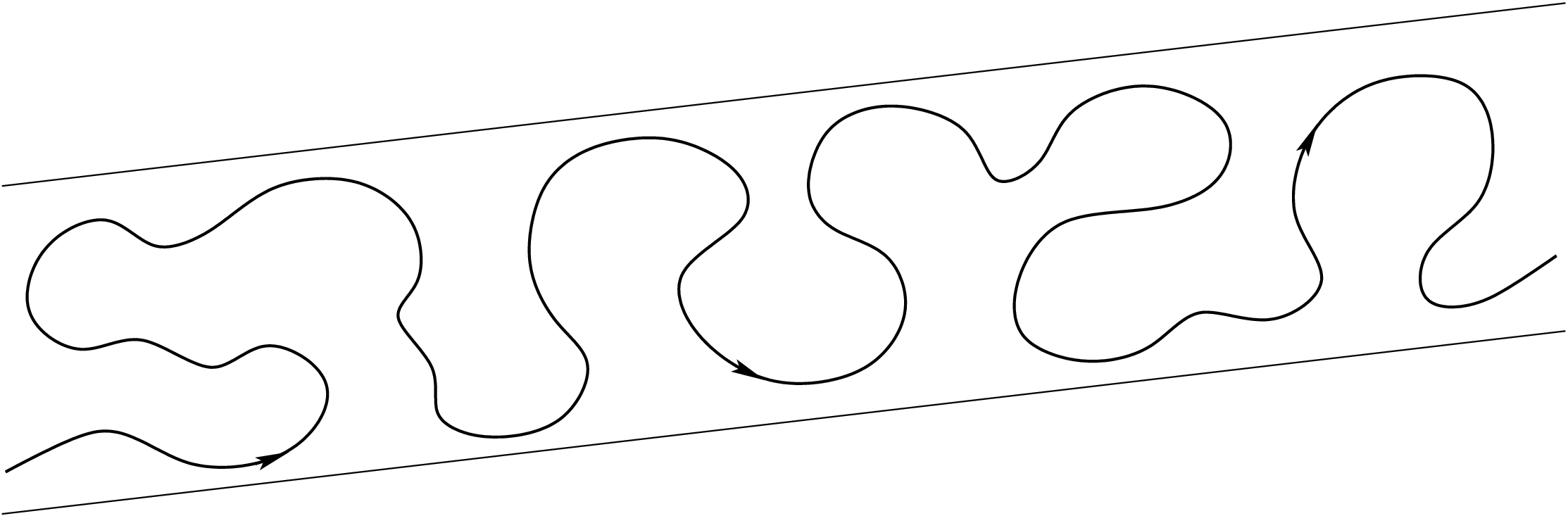}
\end{center}
\caption{The shape of a stable open trajectory of system 
(\ref{MFSyst}) in a plane orthogonal to $\, {\bf B} \, $ 
(schematically).}  
\label{Fig4}
\end{figure}

 The main type of open trajectories of system (\ref{MFSyst}) 
are stable open trajectories, i.e. trajectories that are 
preserved and do not change their shape significantly under 
small rotations of $\, {\bf B} \, $, as well as variations 
in the level $\, \epsilon_{F} \, $. According to 
\cite{zorich1,dynn1992,dynn1}, stable open trajectories of 
(\ref{MFSyst}) have remarkable properties. Namely:

\vspace{1mm}

\noindent
1) Each stable open trajectory of (\ref{MFSyst}) lies in a 
straight strip of finite width (in a plane orthogonal to 
$\, {\bf B}$), passing through it (Fig. \ref{Fig4}).

\vspace{1mm}

\noindent
2) The mean direction of stable open trajectories in 
$\, {\bf p}$ - space is given by the intersection of the 
plane orthogonal to $\, {\bf B} \, $ and some integer 
(generated by two reciprocal lattice vectors) 
plane $\, \Gamma \, $, which is unchanged under small 
rotations of $\, {\bf B} \, $.

\vspace{1mm}

 The contribution of stable open trajectories to the tensor 
$\, \sigma^{kl} ({\bf B}) \, $ is also strongly anisotropic 
in the plane orthogonal to $\, {\bf B} \, $ and is described 
by the formula (\ref{PeriodicTr}) in the leading order at 
$\, B \rightarrow \infty \, $. This property is particularly 
convenient for the experimental study of such trajectories and, 
in particular, played an important role in the introduction of 
topological numbers observable in the conductivity of normal 
metals in the work \cite{PismaZhETF}.

 The emergence of stable open trajectories on the Fermi 
surface corresponds to two-dimensional regions 
$\, \Omega_{\alpha} \subset \mathbb{S}^{2} \, $ 
(Stability Zones) on the angular diagram, each of which 
corresponds to its own integer plane $\, \Gamma_{\alpha} \, $. 
As is easy to see, the emergence of stable open trajectories 
is possible only on sufficiently complex Fermi surfaces 
(like Fig. \ref{Fig1}).

 Angular diagrams containing Stability Zones 
$\, \Omega_{\alpha} \, $ arise in the energy range
$$\epsilon_{F} \, \in \, \left( \epsilon^{\cal A}_{1} , \, 
\epsilon^{\cal A}_{2} \right) \,\,\, , $$
narrower than the full range of values of 
$\, \epsilon ({\bf p}) \, $:
$$\epsilon_{\min} \,\,\, \leq \,\,\, \epsilon^{\cal A}_{1}
\,\,\, < \,\,\, \epsilon^{\cal A}_{2} \,\,\, \leq \,\,\,
\epsilon_{\max} $$

 We will call such diagrams complex conductivity diagrams.
 
\vspace{1mm} 
 
 In fact, the interval 
$\, \left( \epsilon^{\cal A}_{1} , \, 
\epsilon^{\cal A}_{2} \right) \, $ 
has an additional structure associated, among other things, 
with the behavior of Hall conductivity in strong magnetic fields 
(\cite{SecBound,UltraCompl}). Namely, in the general case, the interval 
$\, \left( \epsilon^{\cal A}_{1} , \, 
\epsilon^{\cal A}_{2} \right) \, $ 
can be divided into 3 intervals
$$\left( \epsilon^{\cal A}_{1} , \, \epsilon^{\cal A}_{2} \right)
\,\,\, = \,\,\, 
\left( \epsilon^{\cal A}_{1} , \, \epsilon^{\cal B}_{1} \right) \cup
\left[ \epsilon^{\cal B}_{1} , \, \epsilon^{\cal B}_{2} \right] \cup
\left( \epsilon^{\cal B}_{2} , \, \epsilon^{\cal A}_{2} \right) 
\,\,\, ,  $$
corresponding to different types of complex angular diagrams.

\begin{figure*}[t]
\begin{center}
\includegraphics[width=\linewidth]{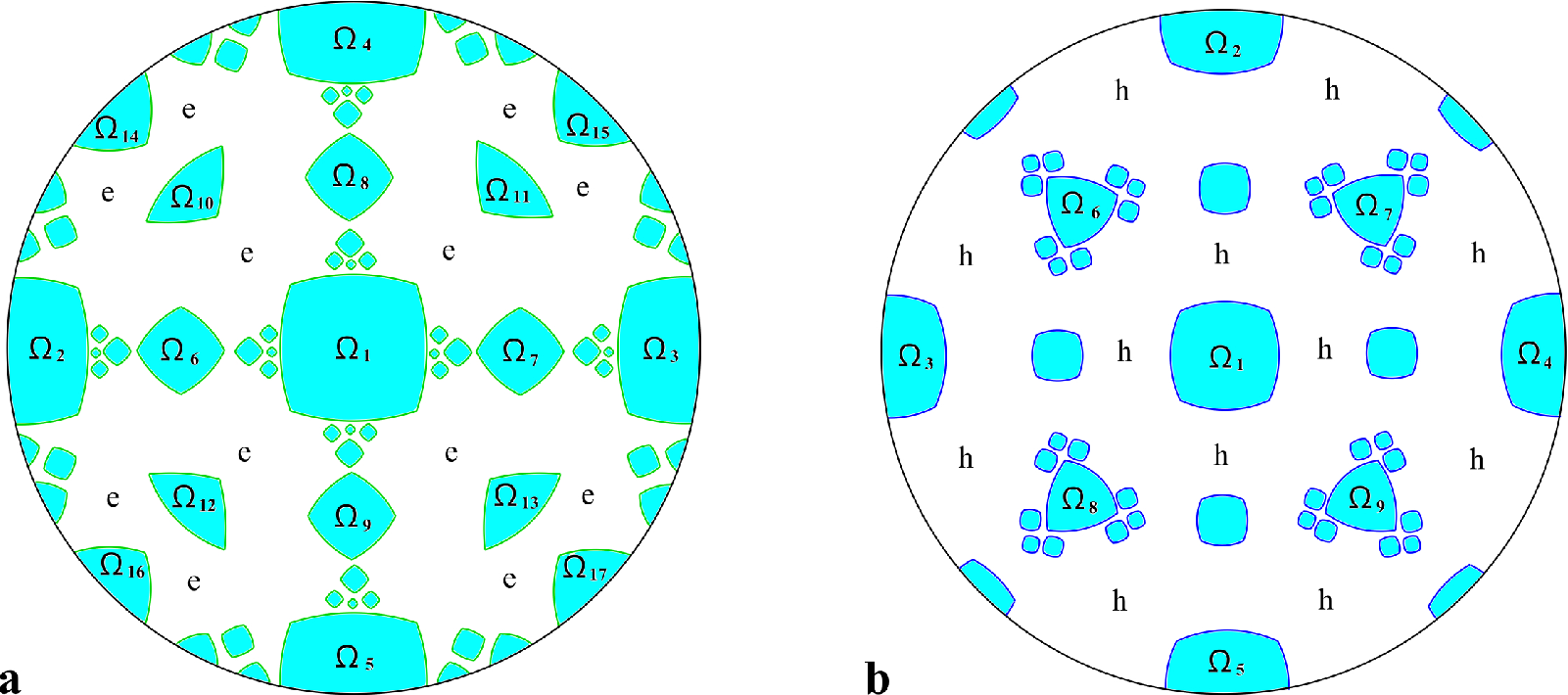}
\end{center}
\caption{(a) Complex angular diagram of type $A_{-}$ (schematically).
(b) Complex angular diagram of type $A_{+}$ (schematically).
(The signs $\, e \, $ and $\, h \, $ denote the type of Hall 
conductivity in regions where only closed trajectories are 
present on the Fermi surface.)}  
\label{Fig5}
\end{figure*}

 For values
$\, \epsilon_{F} \in
\left( \epsilon^{\cal A}_{1} , \, \epsilon^{\cal B}_{1} \right) \, $ 
the conductivity diagrams (diagrams of type $A_{-}$) contain a finite 
number of Stability Zones (Fig. \ref{Fig5}, a). Most of the diagram, 
as a rule, is filled with directions $\, {\bf B} \, $, corresponding 
to the presence of only closed trajectories on the Fermi surface. 
For all such directions $\, {\bf B} \, $ the Hall conductivity 
is of the electron type, and one can use the relation
\begin{equation}
\label{AminusHall}
\sigma^{xy} \,\,\, = \,\,\, - \sigma^{yx} \,\,\, = \,\,\,  
{e c \over B} \,\, {2 V_{-} \over (2 \pi \hbar)^{3}} \,\,\, ,  
\end{equation}
where $\, V_{-} \, $ is the volume of the region
$\, \epsilon ({\bf p}) < \epsilon_{F} \, $ in the torus 
$\, \mathbb{T}^{3} = \mathbb{R}^{3} \big/ L^{*} \, $. 
It can be seen that the Hall conductivity has the same value 
(for a given value of $B$) in this part of the angular diagram.

\vspace{1mm}

 Similarly, for 
$\, \epsilon_{F} \in
\left( \epsilon^{\cal B}_{2} , \, \epsilon^{\cal A}_{2} \right) \, $
the conductivity diagrams (diagrams of type $A_{+}$) contain a finite 
number of Stability Zones (Fig. \ref{Fig5}, b). Most of the diagram 
also corresponds to the presence of only closed trajectories on the 
Fermi surface. The Hall conductivity in this part is of the hole type 
and is given by the relation
\begin{equation}
\label{AplusHall}
\sigma^{xy} \,\,\, = \,\,\, - \sigma^{yx} \,\,\, = \,\,\, - \,\, 
{e c \over B} \,\, {2 V_{+} \over (2 \pi \hbar)^{3}} \,\,\, ,  
\end{equation}
where $\, V_{+} \, $ is the volume of the region 
$\, \epsilon ({\bf p}) > \epsilon_{F} \, $ in the torus
$\, \mathbb{T}^{3} = \mathbb{R}^{3} \big/ L^{*} \, $.

\vspace{1mm}

\begin{figure*}[t]
\begin{center}
\includegraphics[width=\linewidth]{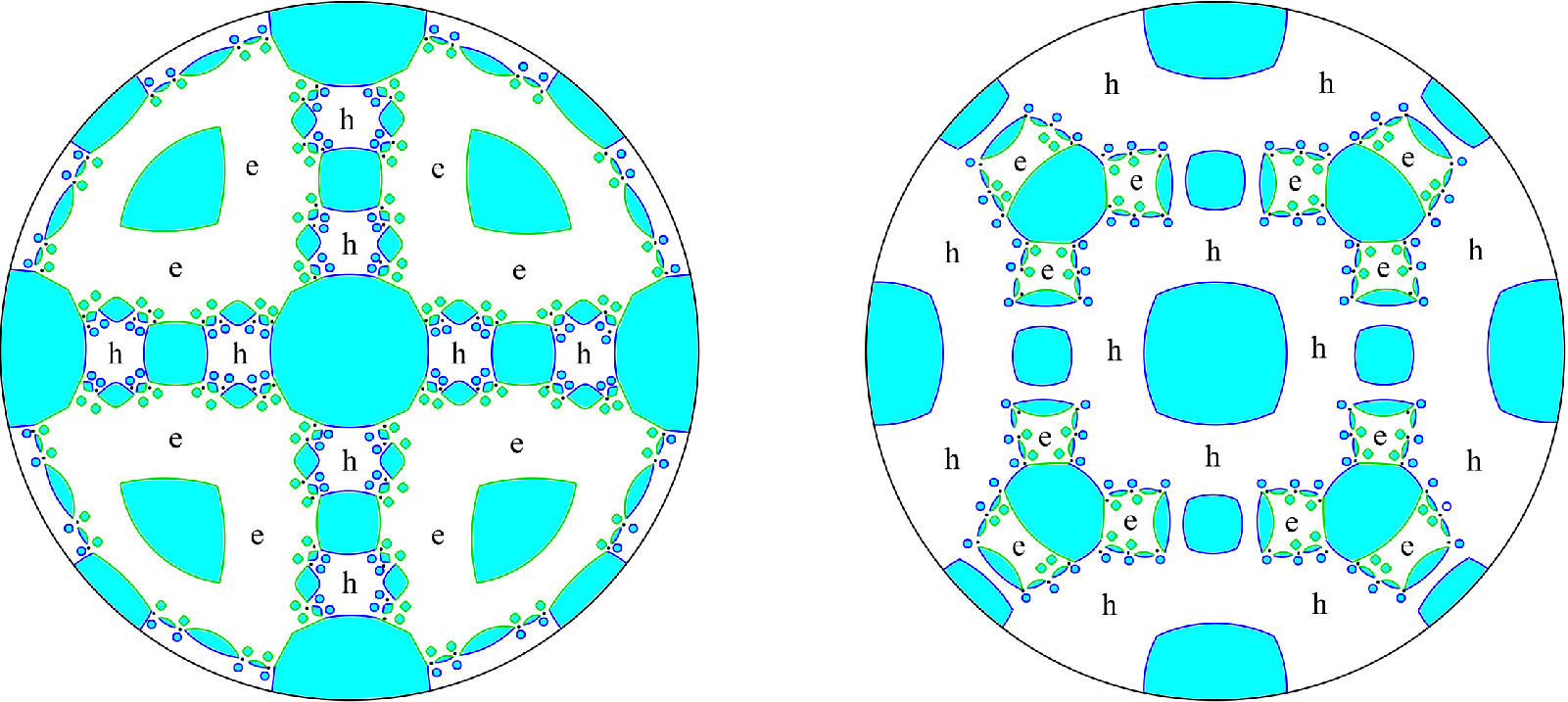}
\end{center}
\caption{Type B diagrams (schematically).
The signs $\, e \, $ and $\, h \, $ denote the type of Hall 
conductivity in regions where only closed trajectories are 
present on the Fermi surface.}  
\label{Fig6}
\end{figure*}

 The diagrams of type $A_{-}$ and $A_{+}$ are separated by even 
more complex diagrams (diagrams of type B), which we will call 
ultra-complex conductivity diagrams (Fig. \ref{Fig6}).

In the generic case, diagrams of type B arise in a finite 
energy interval 
$\, \left[ \epsilon^{\cal B}_{1} , \epsilon^{\cal B}_{2} \right] \, $ 
(\cite{SecBound,UltraCompl}), separating the intervals 
$\, \left( \epsilon^{\cal A}_{1} , \epsilon^{\cal B}_{1} \right) \, $ 
and $\, \left( \epsilon^{\cal B}_{2} , \epsilon^{\cal A}_{2} \right) \, $. 
Generic type B diagrams contain an infinite number of zones 
$\, \Omega_{\alpha} \, $, and among the regions corresponding 
to the presence of only closed trajectories on the Fermi surface, 
there are both regions of electron and hole Hall conductivity 
(Fig. \ref{Fig6}). As before, in regions of the first type, 
one can use the formula (\ref{AminusHall}), and in regions of 
the second type, the formula (\ref{AplusHall}).

\vspace{1mm}

\begin{figure*}[t]
\begin{center}
\includegraphics[width=\linewidth]{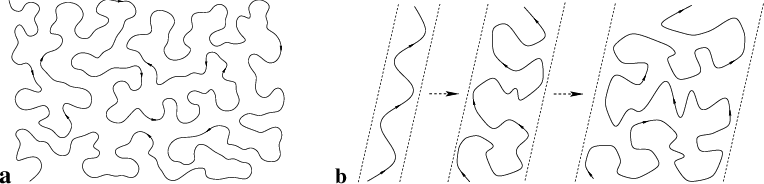}
\end{center}
\caption{(a) Chaotic trajectory of the Dynnikov type in a plane 
orthogonal to $\, {\bf B} \, $ (schematically). 
(b) Increasing complexity of the shape of stable open 
trajectories with decreasing sizes of the 
Zones $\, \Omega_{\alpha} \, $. }  
\label{Fig7}
\end{figure*}

 Accumulation points of small Zones $\, \Omega_{\alpha} \, $ 
typically correspond to the emergence of complex ``chaotic'' 
trajectories (of the Tsarev or Dynnikov type) on the Fermi surface. 
The most complex of these (Dynnikov-type trajectories) have a very 
complex shape, ``wandering everywhere'' in planes orthogonal 
to $\, {\bf B} \, $ (Fig. \ref{Fig7}, a).

 The complex shape of such trajectories is also reflected in 
their contribution to the conductivity tensor in the limit 
$\, B \rightarrow \infty \, $ (suppression of conductivity along 
the direction of $\, {\bf B} \, $, the emergence of fractional 
powers of $\, \omega_{B} \tau \, $ in the components 
$\, \sigma^{kl} ({\bf B}) \, $, etc. \cite{ZhETF1997,TrMIAN}). 
We also note here that numerous properties of chaotic trajectories 
in general are of great interest and are actively studied at 
the present time (see \cite{zorich3,DeLeo1,DeLeo2,DeLeo3,DeLeoDynnikov1,
Dynnikov2008,DeLeoDynnikov2,Skripchenko1,Skripchenko2,DynnSkrip1,
DynnSkrip2,AvilaHubSkrip1,AvilaHubSkrip2,DynHubSkrip,DynMalNovUMN}). 

 It can also be noted that the diminishing of the Zones 
$\, \Omega_{\alpha} \, $ is accompanied by a complication of 
the shape of stable open trajectories (Fig. \ref{Fig7}, b) 
and a complication of their contribution to the tensor 
$\, \sigma^{kl} ({\bf B}) \, $, gradually transferring them 
to the ``chaotic'' regime.

\vspace{1mm}

 Although the occurrence of type B diagrams is a general 
property of generic dispersion laws, they have not yet been 
observed experimentally. In our opinion, this is due to the 
extreme narrowness of the interval 
$\, \left[ \epsilon^{\cal B}_{1} , \epsilon^{\cal B}_{2} \right] \, $ 
for real dispersion laws. In this paper, we estimate the position 
of the interval 
$\, \left[ \epsilon^{\cal B}_{1} , \epsilon^{\cal B}_{2} \right] \, $ 
and its width for analytical relations $\, \epsilon ({\bf p}) \, $ 
based on the tight-binding approximation for crystals with cubic 
symmetry.

 The paper \cite{TightBind} contains estimates for the position of 
the interval 
$\, \left[ \epsilon^{\cal B}_{1} , \epsilon^{\cal B}_{2} \right] \, $ 
for the simple, face-centered, and body-centered cubic lattices in the 
leading order of the tight-binding approximation. It should be noted, 
however, that the leading order of the tight-binding approximation 
does not allow one to estimate the width of the interval 
$\, \left[ \epsilon^{\cal B}_{1} , \epsilon^{\cal B}_{2} \right] \, $ 
for the simple and body-centered lattices and determines only 
its position ($\epsilon_{F} \simeq 0$). The reason for this is that 
the leading order of the tight-binding approximation  
gives here non-generic dispersion relations $\, \epsilon ({\bf p}) \, $  
with the intervals 
$\, \left[ \epsilon^{\cal B}_{1} , \, \epsilon^{\cal B}_{2} \right] \, $ 
contracted to a point 
($\epsilon^{\cal B}_{1} = \epsilon^{\cal B}_{2}$).

 The property described above is, in fact, inherent in all 
dispersion relations $\, \epsilon ({\bf p}) \, $ that contain only odd 
Fourier harmonics, such that
$$\epsilon  \left( {\bf p} + {1 \over 2} \left( 
{\bf a}_{1} + {\bf a}_{2} + {\bf a}_{3} \right) \right) 
\,\, = \,\, - \, \epsilon ({\bf p}) $$

 In particular, this applies to the dispersion relation
$$ \epsilon ({\bf p}) \,\,\, \sim \,\,\, \cos {p_{x} l \over \hbar} 
\,\, + \,\, \cos {p_{y} l \over \hbar} \,\, + \,\, 
\cos {p_{z} l \over \hbar} \,\,\, , $$
arising in the leading order of the tight-binding approximation 
for the simple cubic lattice. The Fermi surface
$$\cos {p_{x} l \over \hbar} \,\, + \,\, \cos {p_{y} l \over \hbar} 
\,\, + \,\, \cos {p_{z} l \over \hbar} \,\,\, = \,\,\, 0 $$
contains open trajectories of (\ref{MFSyst}) for any direction 
of $\, {\bf B} \, $ (\cite{DynnBuDA}), and the corresponding 
angular diagram has the most complex form (being completely 
filled with Zones $\, \Omega_{\alpha} \, $, as well as ``chaotic'' 
and ``special rational'' directions of $\, {\bf B} \, $ 
\cite{DynnBuDA,DeLeo2017,DynMalNovUMN}). For 
$\, \epsilon_{F} < 0 \, $ the angular conductivity diagrams 
are of type $\, A_{-} \, $ (or belong to the class of simple ones), 
and for $\, \epsilon_{F} > 0 \, $ - of type $\, A_{+} \, $ 
(or belong to the class of simple ones).

 A similar situation also arises in the case of a body-centered
cubic lattice, where the dispersion relation in the leading order 
has the form
$$ \epsilon ({\bf p}) \,\,\, \sim \,\,\, \cos {p_{x} l \over \hbar} 
\, \cos {p_{y} l \over \hbar} \, \cos {p_{z} l \over \hbar} $$

 In this paper, we estimate the width of the interval 
$\, \left[ \epsilon^{\cal B}_{1} , \, \epsilon^{\cal B}_{2} \right] \, $ 
for the simple and body-centered cubic lattices, taking into account 
higher-order corrections in the tight-binding approximation. The 
corrections to the leading term are given here by the higher 
Fourier harmonics determined by the lattice geometry. From the 
general point of view, the corresponding dispersion laws are 
a special case of analytic dispersion relations, which are also 
widely considered in the literature (see, e.g., \cite{lifpes1,lifpes2,etm}). 
In the next chapter, we briefly describe a simplified method for estimating 
the position of the interval 
$\, \left[ \epsilon^{\cal B}_{1} , \, \epsilon^{\cal B}_{2} \right] \, $ 
and also carry out this estimate for the simple cubic lattice. 
In Chapter 3, we will estimate the position of 
$\, \left[ \epsilon^{\cal B}_{1} , \, \epsilon^{\cal B}_{2} \right] \, $ 
for the body-centered lattice, where the Fermi surfaces have a more complex 
geometry.

\section{Methods for estimating the interval 
$\, \left[ \epsilon^{\cal B}_{1} , \, \epsilon^{\cal B}_{2} \right] \, $ 
and the case of the simple cubic lattice}
\setcounter{equation}{0}

\begin{figure*}[t]
\begin{center}
\includegraphics[width=\linewidth]{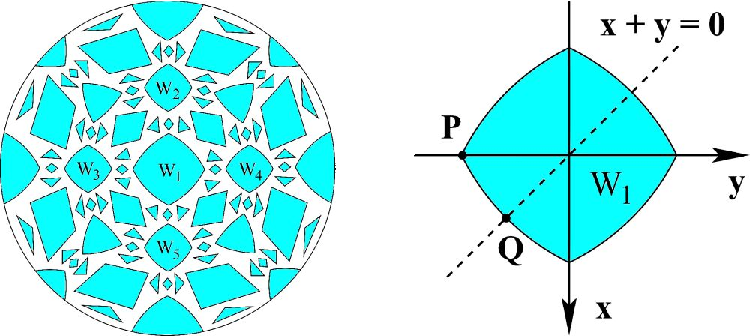}
\end{center}
\caption{(a) Zones $\, W_{\alpha} \, $ on the ``full'' angular 
diagram for a dispersion relation $\, \epsilon ({\bf p}) \, $. 
(b) ``Symmetric'' points $P$ and $Q$ on the boundary of the 
symmetric Zone $\, W_{1} \, $.}  
\label{Fig8}
\end{figure*}

 To find the interval 
$\, \left[ \epsilon^{\cal B}_{1} , \, \epsilon^{\cal B}_{2} \right] \, $, 
it is necessary to study the system (\ref{MFSyst}) on the Fermi surfaces
$\, \epsilon ({\bf p}) = \epsilon_{F} \, $ for various values of
$\, \epsilon_{F} \, $. In fact, angular diagrams similar to those for 
the Fermi surfaces can be constructed for the entire dispersion law 
(\cite{dynn3}). Such angular diagrams contain Stability Zones 
$\, W_{\alpha} \, $, which now form an everywhere dense set on the 
sphere $\, \mathbb{S}^{2} \, $ (Fig. \ref{Fig8}, a). Each of the Zones 
$\, \Omega_{\alpha} \, $ is a subregion of some Zone $\, W_{\alpha} \, $, 
corresponding to the same integer plane $\, \Gamma_{\alpha} \, $.

\begin{figure*}[t]
\begin{center}
\includegraphics[width=\linewidth]{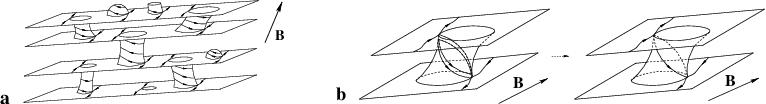}
\end{center}
\caption{(a) Complex Fermi surface divided by closed trajectories 
into carriers of open trajectories.
(b) Vanishing of the height of a cylinder of closed trajectories at 
the boundary of a Zone $\, \Omega_{\alpha} \, $ or $\, W_{\alpha} \, $. }  
\label{Fig9}
\end{figure*}

  For each $\, {\bf n} = {\bf B}/B \in \Omega_{\alpha} \, $ 
system (\ref{MFSyst}) has a certain structure on the corresponding 
Fermi surface (\cite{zorich1,dynn1,dynn3}). Namely, the Fermi 
surface contains a certain number of cylinders of closed trajectories 
dividing it into ``carriers of open trajectories'', which, in turn, 
are periodically deformed integer planes (with holes) in 
$\, {\bf p}$ - space (Fig. \ref{Fig9}, a). At the boundary of a
Zone $\, \Omega_{\alpha} \, $ the described structure of trajectories 
is destroyed due to the vanishing of the height of one of the 
cylinders of closed trajectories separating the carriers of open 
trajectories (Fig. \ref{Fig9}, b).

 For $\, {\bf n} \in W_{\alpha} \, $ a similar structure of 
system (\ref{MFSyst}) is preserved in a certain interval
$$\epsilon_{F} \,\,\, \in \,\,\, \left[ 
\widetilde{\epsilon}_{1} ({\bf n}) , \, 
\widetilde{\epsilon}_{2} ({\bf n}) \right] \,\,\, , $$
where $\, \widetilde{\epsilon}_{1} ({\bf n}) \, $ and 
$\, \widetilde{\epsilon}_{2} ({\bf n}) \, $ are some continuous 
functions on the sphere $\, \mathbb{S}^{2} \, $ (\cite{dynn3}), 
such that
$$\widetilde{\epsilon}_{1} ({\bf n}) \,\,\, \leq \,\,\,
\widetilde{\epsilon}_{2} ({\bf n}) $$
(i.e. at every point of the sphere).

 The boundary of a Zone $\, W_{\alpha} \, $ is determined 
by the disappearance of at least two cylinders of closed 
trajectories (electron and hole types) separating the carriers 
of open trajectories. This condition generally determines both 
the position of the boundary of $\, W_{\alpha} \, $ and the 
value of the function
$$\widetilde{\epsilon}_{0} ({\bf n}) \,\,\, = \,\,\, 
\widetilde{\epsilon}_{1} ({\bf n}) \,\,\, = \,\,\,
\widetilde{\epsilon}_{2} ({\bf n}) $$
for $\, {\bf n} \in \partial W_{\alpha} \, $ (\cite{dynn3}).

\vspace{1mm}
 
 According to \cite{SecBound,UltraCompl}, the quantities
$\, \epsilon^{\cal B}_{1} \, $ and $\, \epsilon^{\cal B}_{2} \, $
satisfy the relations
\begin{equation}
\label{epsilonB12}
\epsilon^{\cal B}_{1} \,\,\, = \,\,\, \min_{\mathbb{S}^{2}} \, 
\widetilde{\epsilon}_{2} ({\bf n}) \quad , \quad \quad
\epsilon^{\cal B}_{2} \,\,\, = \,\,\, \max_{\mathbb{S}^{2}} \, 
\widetilde{\epsilon}_{1} ({\bf n}) \,\,\, , 
\end{equation}

 Globally, in the generic case, we have the relation
$$\max_{\mathbb{S}^{2}} \, \widetilde{\epsilon}_{1} ({\bf n}) 
\,\,\, > \,\,\, 
\min_{\mathbb{S}^{2}} \, \widetilde{\epsilon}_{2} ({\bf n}) $$

\vspace{1mm}

 Finding the functions $\, \widetilde{\epsilon}_{1} ({\bf n}) \, $ 
and $\, \widetilde{\epsilon}_{2} ({\bf n}) \, $ is in general a rather 
complex computational problem. However, for an approximate calculation 
of the width of the interval 
$\, \left[ \epsilon^{\cal B}_{1} , \, \epsilon^{\cal B}_{2} \right] \, $, 
one can use special methods that give a good (and in some cases accurate) 
estimate of its position. Here we use the method proposed in \cite{TightBind} 
and suitable for many relations $\, \epsilon ({\bf p}) \, $ that 
have a fairly rich symmetry.
 
 Namely, we first replace the expressions (\ref{epsilonB12}) 
with the expressions
\begin{equation}
\label{BoundAppr}
\epsilon^{\cal B}_{1} \,\,\, = \,\,\, 
\min_{\cup \partial W_{\alpha}} \, \widetilde{\epsilon}_{0} ({\bf n})
\,\,\, , \quad
\epsilon^{\cal B}_{2} \,\,\, = \,\,\, 
\max_{\cup \partial W_{\alpha}} \, \widetilde{\epsilon}_{0} ({\bf n})
\end{equation}
(the minimum and maximum are taken only along the boundaries 
of the Zones $W_{\alpha}$), which give the same values of
$\, \epsilon^{\cal B}_{1} \, $ and $\, \epsilon^{\cal B}_{2} \, $
for the majority of ``physically realistic'' relationships 
$\, \epsilon ({\bf p}) \, $

 Second, to evaluate the expressions (\ref{BoundAppr}), we use 
the values of $\, \widetilde{\epsilon}_{0} ({\bf n}) \, $ at the 
boundaries of the largest Zones $\, W_{\alpha} \, $, where their 
variation is sufficiently large. More precisely, to evaluate the 
values of $\, \epsilon^{\cal B}_{1} \, $ and $\, \epsilon^{\cal B}_{1} \, $,
we use the values of $\, \widetilde{\epsilon}_{0} (P) \, $ and
$\, \widetilde{\epsilon}_{0} (Q) \, $ at the ``symmetric'' points 
$\, P \, $ and $\, Q \, $ of the boundary of the symmetric Zone 
$\, W_{1} \, $, shown in Fig. \ref{Fig8}, b. By tracking the 
disappearance of cylinders of closed trajectories on the 
corresponding Fermi surfaces, we can determine the position of 
the points $\, P \, $ and $\, Q \, $, as well as the values of
$\, \widetilde{\epsilon}_{0} (P) \, $ and 
$\, \widetilde{\epsilon}_{0} (Q) \, $.

\vspace{1mm}

 In this section, we estimate the position of the interval 
$\, \left[ \epsilon^{\cal B}_{1} , \epsilon^{\cal B}_{2} \right] \, $ 
in the tight-binding approximation for a simple cubic lattice. 
For simplicity, we will use ``dimensionless'' values of 
$\, \epsilon \, $ and $\, {\bf p} \, $ and denote the components 
$\, {\bf p} \, $ by the usual coordinates $\, {\bf p} = (x, y, z) \, $.

\begin{figure}[t]
\begin{center}
\includegraphics[width=0.8\linewidth]{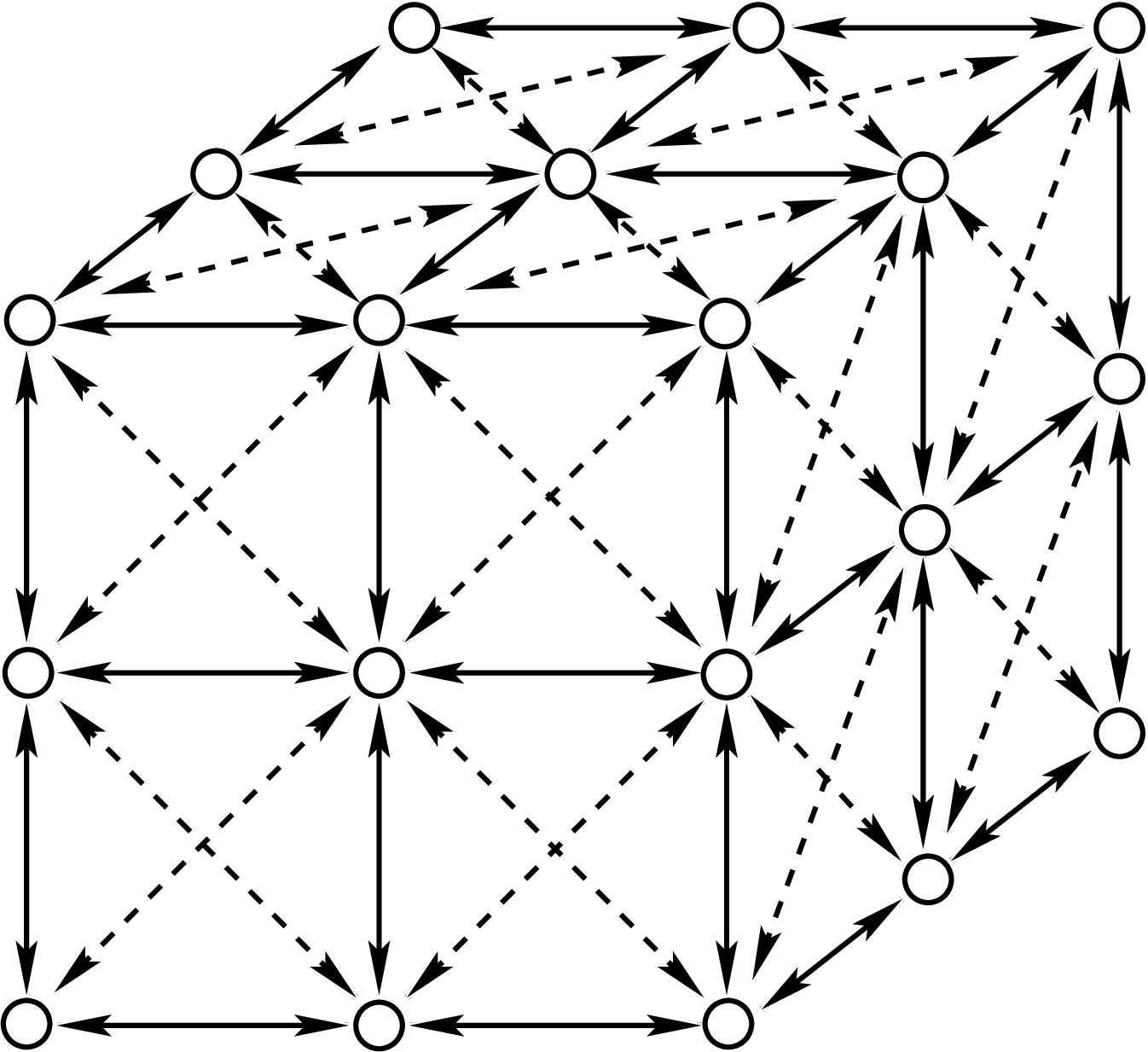}
\end{center}
\caption{Principal and additional jumps between nodes of a simple 
cubic lattice in the tight-binding approximation.}  
\label{Fig10}
\end{figure}

 As a correction to the leading approximation, we consider here 
jumps to sites next to the nearest ones (Fig. \ref{Fig10}). 
The full dispersion relation takes then the form
\begin{multline}
\label{RelSimpleCube}
\epsilon_{\delta} ({\bf p}) \,\,\, = \,\,\,
\cos x \,\, + \,\, \cos y \,\, + \,\, \cos z  \,\, +  \\
+ \,\, \delta \, \Big( \cos \, (x+y) \, + \, \cos \, (x-y) \, + \,
\cos \, (x+z) \, +  \\
+ \, \cos \, (x-z) \, + \cos \, (y+z) \, + \, \cos \, (y-z) 
\Big) \,\,\, =  \\
= \,\,\, \cos x \,\, + \,\, \cos y \,\, + \,\, \cos z  \,\, +  \\
+ \,\, 2 \delta \left( \cos x \, \cos y \, + \, \cos x \, \cos z
\, + \, \cos y \, \cos z  \right)
\end{multline}

 The equations of the Fermi surfaces
\begin{multline*}
\cos x \,\, + \,\, \cos y \,\, + \,\, \cos z  \,\, +  \\
+ \,\, 2 \delta \left( \cos x \, \cos y \, + \, \cos x \, \cos z
\, + \, \cos y \, \cos z  \right)
\,\, =  \,\, \epsilon_{F} 
\end{multline*}
are invariant under the transformation
$$x \,\, \rightarrow \,\, x \, + \, \pi \,\,\, , \quad
y \,\, \rightarrow \,\, y \, + \, \pi \, \,\,, \quad
z \,\, \rightarrow \,\, z \, + \, \pi $$
$$\delta \,\, \rightarrow \,\, - \, \delta \,\,\, , \quad
\epsilon_{F} \,\, \rightarrow \,\, - \, \epsilon_{F} $$

 It can be seen, therefore, that the Fermi surfaces 
$\, \epsilon_{-\delta} ({\bf p}) = \epsilon_{F} \, $ are 
obtained from the surfaces 
$\, \epsilon_{\delta} ({\bf p}) = - \epsilon_{F} \, $ 
by a shift in $\, {\bf p}$ - space. Thus, in the study of 
the trajectories of (\ref{MFSyst}), it is sufficient for us 
to restrict ourselves to the values $\, \delta \geq 0 \, $ 
(the reality of $\, \delta \, $ is a consequence 
of the symmetry of the lattice).

\begin{figure*}[t]
\begin{center}
\includegraphics[width=\linewidth]{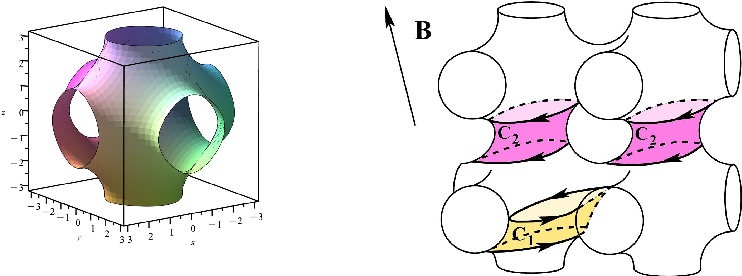}
\end{center}
\caption{The Fermi surfaces
$\, \epsilon_{F} ({\bf p}) = \epsilon_{F} \, $
for sufficiently small values of $\, \delta \, $ 
and $\, \epsilon_{F} \, $ and cylinders of closed
trajectories cutting the Fermi surfaces into carriers
of open trajectories (simple cubic lattice).}
\label{Fig11}
\end{figure*}

 For $\, \delta = 0 \, $ the Fermi surface 
$\, \epsilon ({\bf p}) = 0 \, $ has genus 3 and rank 3 
(Fig. \ref{Fig11}) and this property is preserved for sufficiently 
small values of $\, \delta \, $ and $\, \epsilon_{F} \, $.

 For directions $\, {\bf B} \, $ close to $\, z \, $, 
it is easy to distinguish cylinders $\, C_{1} \, $ 
and $\, C_{2} \, $ (of electron and hole type) that cut these 
surfaces into carriers of open trajectories for 
$\, {\bf n} \in W_{1} \, $ (and 
$\, \epsilon_{F} \in \left[ \widetilde{\epsilon}_{1} ({\bf n}) , \, 
\widetilde{\epsilon}_{2} ({\bf n}) \right]$) (Fig. \ref{Fig11}).

 It is easy to see that the relation (\ref{RelSimpleCube}), 
as well as the Fermi surfaces 
$\, \epsilon ({\bf p}) = \epsilon_{F} \, $, 
are symmetric with respect to the planes
$$\Pi_{n} \,: \quad x \,\,\, = \,\,\, \pi n \,\,\, , 
\quad n \in \mathbb{Z} $$
(as well as the planes $\, y = \pi n \, $, $\, z = \pi n \, $,
$\, n \in \mathbb{Z} $).

 The point $\, P \in \mathbb{S}^{2} \, $ (Fig. \ref{Fig8}, b) 
corresponds to the disappearance of both cylinders 
$\, C_{1} \, $ and $\, C_{2} \, $. According to Fig. \ref{Fig9}, b, 
this corresponds to the emergence of cylinders of zero height 
containing a pair of saddle singular points of system (\ref{MFSyst}). 
As is not difficult to see, by virtue of symmetry, these saddle 
singular points lie in the planes $\, \Pi_{n} \, $.

\begin{figure*}[t]
\begin{center}
\includegraphics[width=\linewidth]{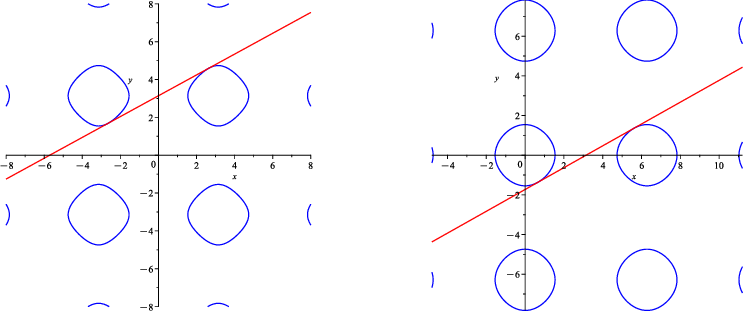}
\end{center}
\caption{Simultaneous tangency of a pair of ovals defined 
by the relation 
$\, \epsilon_{\delta} ({\bf p}) = \epsilon_{F} \, $ 
($\delta = 0.2 , \,\epsilon_{F} = -0.1728265816 $) 
by straight lines $\, \Pi ({\bf B}) \cap \Pi_{n} \, $
in even and odd planes $\, \Pi_{n} \, $ 
(simple cubic lattice).}  
\label{Fig12}
\end{figure*}

 In the planes $\, \Pi_{n} \, $, the disappearance of the 
cylinders $\, C_{1} \, $ and $\, C_{2} \, $ corresponds to the 
simultaneous tangency of a pair of ovals defined by the relation
$$\epsilon ({\bf p}) \Big|_{\Pi_{n}} \,\,\, = \,\,\, \epsilon_{F} 
\,\,\, , $$
by some planes $\, \Pi ({\bf B}) \, $, orthogonal to 
$\, {\bf B} \, $ (or, equivalently, by lines 
$\, \Pi ({\bf B}) \cap \Pi_{n} $). In particular, in even 
planes $\, \Pi_{n} \, $, the lines 
$\, \Pi ({\bf B}) \cap \Pi_{n} \, $ must touch the pairs of 
ovals defined by the relation
\begin{multline*}
\epsilon ({\bf p}) \Big|_{\Pi_{n}} \,\,\, =  \\
= \,\, \left( 1 + 2 \delta \right) (\cos y \, + \, \cos z ) 
\,\, + \,\, 2 \delta \, \cos y \,\, \cos z \,\,\, = \,\,\, 
\epsilon_{F} \, - \, 1 \,\,\, , 
\end{multline*}
and in odd ones
$$\left( 1 - 2 \delta \right) (\cos y \, + \, \cos z ) 
\,\, + \,\, 2 \delta \, \cos y \,\, \cos z \,\,\, = \,\,\, 
\epsilon_{F} \, + \, 1 $$
(Fig. \ref{Fig12}).

 The equality of the slope of the tangents in the even 
and odd planes $\, \Pi_{n} \, $ gives us the required 
dependence of $\, \epsilon_{F} \, $ 
(i.e. $\, \widetilde{\epsilon}_{0} (P) $) on the 
value $\, \delta \, $.

\vspace{1mm}

 Similarly, the point $\, Q \in \mathbb{S}^{2} \, $ 
corresponds to the emergence of zero-height cylinders 
($C_{1}$ and $C_{2}$) containing singular points of the 
system (\ref{MFSyst}). As can be seen from the geometry 
of our Fermi surfaces, each zero-height cylinder now 
contains 4 saddle singular points of (\ref{MFSyst}). 
For topological reasons, the total number of saddle 
singular points on our surfaces is also equal to 4. 
It follows then that each plane $\, \Pi ({\bf B}) \, $ 
containing zero-height cylinders $\, C_{1} \, $ must 
also contain zero-height cylinders $\, C_{2} \, $ 
(adjacent to the same saddles).

 For the value $\, \delta = 0 \, $ we have 
$\, \widetilde{\epsilon}_{0} (Q) = 0 \, $, and the 
plane we need (for example) is the plane
\begin{equation}
\label{PiBplane}
z \,\,\, = \,\,\, (y - x) / 2 
\end{equation}

\begin{figure*}[t]
\begin{center}
\includegraphics[width=\linewidth]{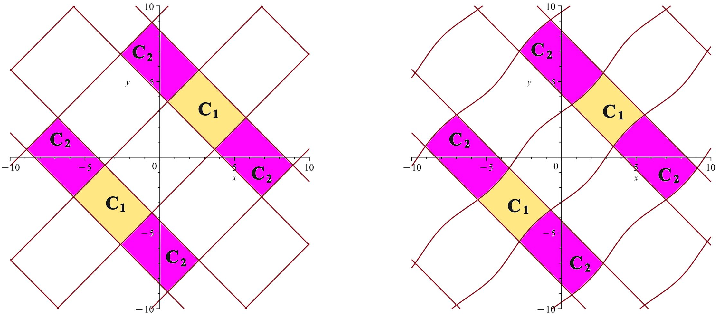}
\end{center}
\caption{The level lines $\, \epsilon_{0} ({\bf p}) = 0 \, $ 
and $\, \epsilon_{\delta} ({\bf p}) = \epsilon_{F} \, $ 
($\delta = 0.2 , \,\epsilon_{F} = - 0.3 $) in the plane 
(\ref{PiBplane}) for the dispersion law (\ref{RelSimpleCube}) 
(simple cubic lattice).}  
\label{Fig13}
\end{figure*}

 As follows from the symmetry of the dispersion law 
(\ref{RelSimpleCube}), for $\, \delta \neq 0 \, $, 
the plane $\, \Pi ({\bf B}) \, $, containing zero-height 
electron-type cylinders, passes through the points 
$\, (- \pi , - \pi , 0 ) \, $ and $\, (\pi , \pi , 0 ) \, $ 
(the centers of the cylinders). Similarly, the plane
$\, \Pi ({\bf B}) \, $, containing zero-height hole-type 
cylinders, passes through the points 
$\, (0 , - 2 \pi , \pi ) \, $ and $\, (2 \pi , 0 , \pi ) \, $. 
Both of these planes obviously coincide with the plane 
(\ref{PiBplane}) for $\, \delta = 0 \, $. In view of the above, 
however, these planes must coincide with each other also for 
$\, \delta \neq 0 \, $. This means that the plane (\ref{PiBplane}) 
remains unchanged for $\, \delta \neq 0 \, $, and only the value 
of $\, \widetilde{\epsilon}_{0} (Q, \delta) \, $, as well as the 
position of the singular points inside this plane, change 
(Fig. \ref{Fig13}). In other words, in our case, the position 
of the point $\, Q \in \mathbb{S}^{2} \, $ remains unchanged 
under variations of $\, \delta \, $.

 Due to the symmetry of the problem, to find the value 
$\, \widetilde{\epsilon}_{0} (Q, \delta) \, $ it is 
sufficient to impose the condition of tangency of the 
plane (\ref{PiBplane}) and the surface 
$\, \epsilon_{\delta} ({\bf p}) = \widetilde{\epsilon}_{0} (Q, \delta) \, $ 
at one of the saddle points (with coordinates $\, (x, y, z) \, $). 
The tangency condition implies collinearity of the gradient of
$\, \epsilon_{\delta} ({\bf p}) \, $
\begin{multline*}
\nabla \epsilon_{\delta} ({\bf p}) \,\,\, = \,\,\, 
\left( 
\begin{array}{c}
- \sin x \left( 1 + 2 \delta \cos y + 2 \delta \cos z \right) \\
- \sin y \left( 1 + 2 \delta \cos x + 2 \delta \cos z \right) \\
- \sin z \left( 1 + 2 \delta \cos x + 2 \delta \cos y \right) 
\end{array}  \right)  \,\,\, =  \\
= \,\,\, \left( 
\begin{array}{c}
- \sin x \left( 1 + 2 \delta \cos y + 2 \delta \cos {y - x \over 2} \right) \\
- \sin y \left( 1 + 2 \delta \cos x + 2 \delta \cos {y - x \over 2} \right) \\
- \sin {y - x \over 2} \, \left( 1 + 2 \delta \cos x + 2 \delta \cos y \right) 
\end{array}  \right)
\end{multline*}
and the vector $\, (1, -1, 2) \, $ at the point $\, (x, y, z) \, $.

\vspace{1mm}

 Solving the system
\begin{multline*}
 \cos x \,\, + \,\, \cos y \,\, + \,\, \cos {y - x \over 2}  \,\, +  \\
+ \,\, 2 \delta \left( \cos x \, \cos y \, + \, 
\left( \cos x \, + \, \cos y \right)
\, \cos {y - x \over 2} \right) \,\,\, =  \\
= \,\,\, \widetilde{\epsilon}_{0} (Q, \delta) 
\end{multline*}
\begin{multline*}
\sin x \left( 1 + 2 \delta \cos y + 2 \delta \cos {y - x \over 2} \right) 
\,\,\, =  \\
= \,\,\, - \, 
\sin y \left( 1 + 2 \delta \cos x + 2 \delta \cos {y - x \over 2} \right)
\end{multline*}
\begin{multline*}
\sin x \left( 1 + 2 \delta \cos y + 2 \delta \cos {y - x \over 2} \right) 
\,\,\, =  \\
= \,\,\,  {1 \over 2} \,\,
\sin {y - x \over 2} \,\, \Big( 1 + 2 \delta \cos x + 2 \delta \cos y \Big) 
\end{multline*}
with respect to $\, x \, $, $\, y \, $ and 
$\, \widetilde{\epsilon}_{0} (Q, \delta) \, $, 
we obtain the dependence we need.

\begin{figure*}[t]
\begin{center}
\includegraphics[width=0.9\linewidth]{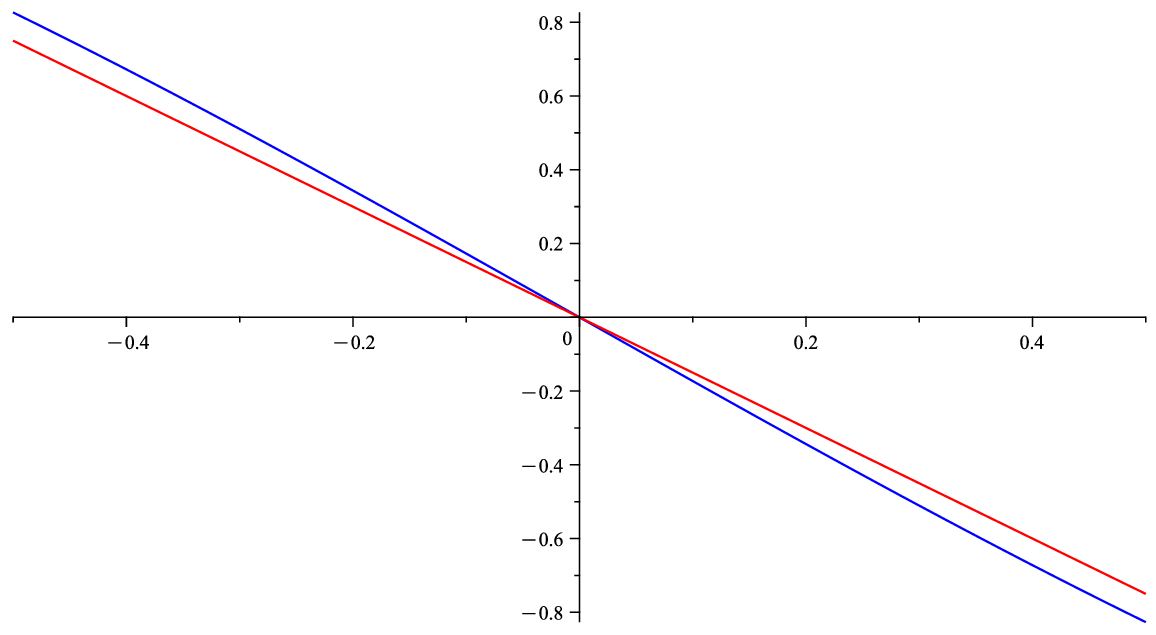}
\end{center}
\caption{The values
$\, \widetilde{\epsilon}_{0} (P, \delta ) \, $ and 
$\, \widetilde{\epsilon}_{0} (Q, \delta ) \, $ for 
the dispersion law (\ref{RelSimpleCube}) in the 
interval $\, - 0.5 < \delta < 0.5 \, $ 
(simple cubic lattice).
}
\label{Fig14}
\end{figure*}

 The values of the functions $\, \widetilde{\epsilon}_{0} (P) \, $
and $\, \widetilde{\epsilon}_{0} (Q) \, $ are presented in 
Fig. \ref{Fig14}. Everywhere in the interval 
$\, 0 < \delta < 0.5 \, $ we have the relation 
$\, \widetilde{\epsilon}_{0} (P) < \widetilde{\epsilon}_{0} (Q) \, $ 
(and $\, \widetilde{\epsilon}_{0} (P) > \widetilde{\epsilon}_{0} (Q) \, $ 
in the interval $\, - 0.5 < \delta < 0 $). As is easy to see, near the 
value $\, \delta = 0.5 \, $ the width of the interval 
$$\left[ \epsilon^{\cal B}_{1} , \, \epsilon^{\cal B}_{2} 
\right] \,\,\, \simeq \,\,\,
\left[ \widetilde{\epsilon}_{0} (P) , \,  
\widetilde{\epsilon}_{0} (Q) \right] $$
is of the order of $ \, 1.5 \% \, $ of the conduction band width.

 For $\, \delta \simeq 0.5 \, $, the topology of the 
Fermi surfaces near the interval 
$\, \left[ \widetilde{\epsilon}_{0} (P) , \, 
\widetilde{\epsilon}_{0} (Q) \right] \, $ changes. 
Specifically, the Fermi surfaces acquire additional 
(very thin) ``handles'' and their genus increases to 7. 
For $\, \delta > 0.5 \, $, therefore, the interval 
$\, \left[ \widetilde{\epsilon}_{0} (P) , \, 
\widetilde{\epsilon}_{0} (Q) \right] \, $ must be 
estimated based on the new geometry of the Fermi 
surfaces. In this paper, we consider the higher terms 
in (\ref{RelSimpleCube}) as corrections to the main 
approximation, so we restrict ourselves here to 
the range of values $\, - 0.5 < \delta < 0.5 \, $
for the simple cubic lattice.

\section{Body-centered cubic lattice}
\setcounter{equation}{0}

\begin{figure}[t]
\begin{center}
\includegraphics[width=0.9\linewidth]{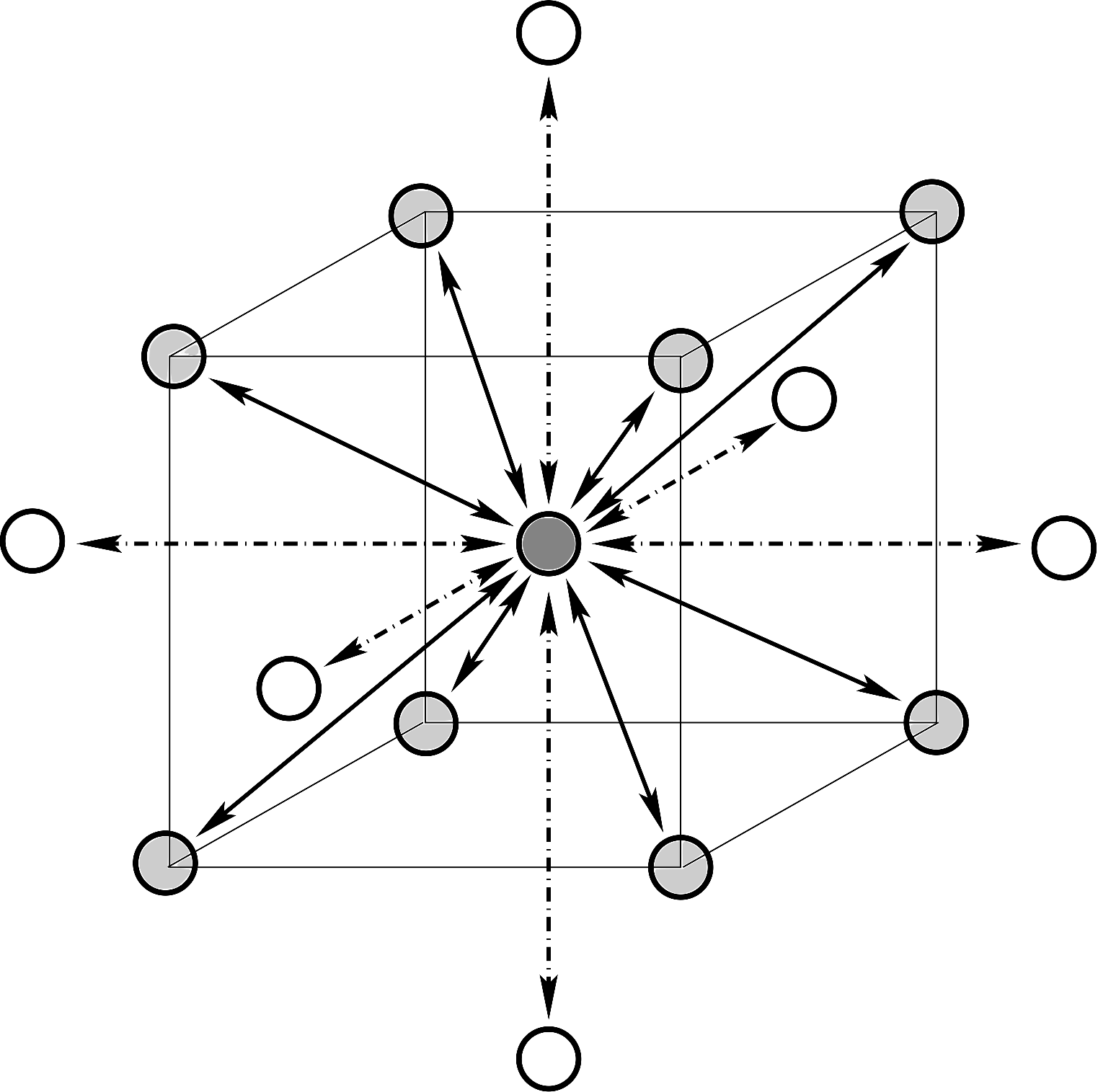}
\end{center}
\caption{Main and additional jumps in the body-centered 
cubic lattice.}  
\label{Fig15}
\end{figure}

 As in the previous section, we will consider here the 
approximation based on jumps between the nearest and next 
lattice nodes (Fig. \ref{Fig15}). In the ``dimensionless'' 
quantities $\, \epsilon \, $ and $\, {\bf p} \, $, it will 
be convenient to use the following normalization
\begin{multline}
\label{VolCentrDispRel}
\epsilon ({\bf p}) \,\,\, = \,\,\, {1 \over 4} \Big[ 
\cos (x + y + z) \, + \, \cos (x + y - z) \, +  \\
+ \, \cos (x - y + z) \, + \, \cos (-x + y + z) \, +  \\
+ \, \, \delta \cos 2 x \, + \, 
\delta \cos 2 y \, + \, \delta \cos 2 z \Big] \,\,\, =  \\
= \,\,\, \cos x \, \cdot \, \cos y \, \cdot \, \cos z \,\, + \,\, 
{\delta \over 4} \Big( \cos 2 x \, + \, \cos 2 y \, + \, 
\cos 2 z \Big)  
\end{multline}

 Here it is also easy to see that the shift
$$x \,\, \rightarrow \,\, x \, + \, \pi \,\,\, , \quad
y \,\, \rightarrow \,\, y \, + \, \pi \, \,\,, \quad
z \,\, \rightarrow \,\, z \, + \, \pi $$
transforms the surfaces
$\, \epsilon_{-\delta} ({\bf p}) = \epsilon_{F} \, $
into the surfaces
$\, \epsilon_{\delta} ({\bf p}) = - \epsilon_{F} \, $.
Thus, as before, it is sufficient for us to study here 
the case $\, \delta \geq 0 \, $ and use the same picture of 
trajectories for $\, \delta \leq 0 \, $ with the replacement 
$\, \epsilon_{F} \rightarrow - \epsilon_{F} \, $.

\begin{figure*}[t]
\begin{center}
\includegraphics[width=\linewidth]{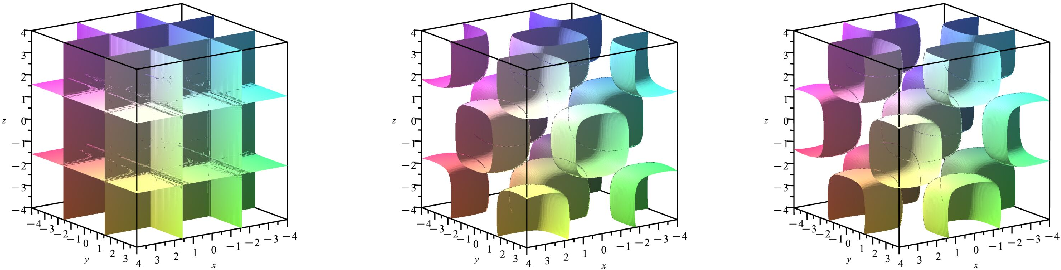}
\end{center}
\caption{The Fermi surfaces (\ref{OsobFermSurf}) for 
$\, \epsilon_{F} = 0 \, $, $\, \epsilon_{F} < 0 \, $ 
and $\, \epsilon_{F} > 0 \, $. 
}
\vspace{0.5cm}
\label{Fig16}
\end{figure*}

 The zeroth-order dispersion relation
$$ \epsilon ({\bf p}) \,\,\, = \,\,\, 
\cos x \, \cdot \, \cos y \, \cdot \, \cos z $$
has here special properties. Namely, an extended Fermi 
surface arises here only for the value 
$\, \epsilon_{F} = 0 \, $, 
for all $\, \epsilon_{F} < 0 \, $ and 
$\, \epsilon_{F} > 0 \, $ the Fermi surfaces
\begin{equation}
\label{OsobFermSurf}
\cos x \, \cdot \, \cos y \, \cdot \, \cos z 
\,\,\, = \,\,\, \epsilon_{F}
\end{equation}
are compact (Fig. \ref{Fig16}).

 It is easy to see that both for $\, \epsilon_{F} < 0 \, $ 
and for $\, \epsilon_{F} > 0 \, $ all trajectories of 
(\ref{MFSyst}) are closed for all directions of 
$\, {\bf B} \, $ (and are of electron and hole type, 
respectively). As a consequence, we have here in the 
zeroth approximation
$$\epsilon^{\cal A}_{1} \,\,\, = \,\,\, 
\epsilon^{\cal B}_{1} \,\,\, = \,\,\, 
\epsilon^{\cal B}_{2} \,\,\, = \,\,\, 
\epsilon^{\cal A}_{2} \,\,\, = \,\,\, 0 $$

 Thus, the emergence of the second term in 
(\ref{VolCentrDispRel}) leads here not only to a non-zero 
width of the interval
$\, \left[ \epsilon^{\cal B}_{1} , \, 
\epsilon^{\cal B}_{2} \right] \, $, but also of the interval 
$\, \left( \epsilon^{\cal A}_{1} , \, 
\epsilon^{\cal A}_{2} \right) \, $.

 For $\, \delta > 0 \, $ the dispersion relation 
(\ref{VolCentrDispRel}) becomes a generic relation. 
As is not difficult to show, extended Fermi surfaces arise 
here in the interval
$$\epsilon_{F} \,\,\, \in \,\,\, \left( 
\epsilon^{\cal A}_{1} , \, \epsilon^{\cal A}_{2} \right)
\,\,\, = \,\,\,   \left(
- {3 \delta \over 4} + {\delta^{3} \over 2} \, , \,\, 
- {\delta \over 4} \right) $$
(and symmetrically for $\, \delta < 0 $).

\begin{figure*}[t]
\begin{center}
\includegraphics[width=\linewidth]{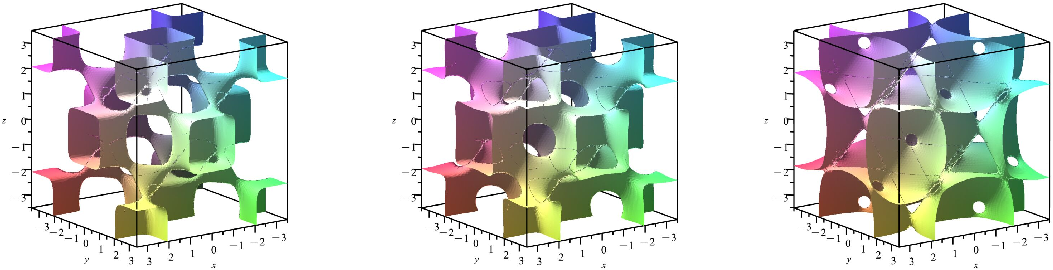}
\end{center}
\caption{The Fermi surfaces 
$\, \epsilon_{\delta} ({\bf p}) = \epsilon_{F} \, 
$ for $\, \epsilon_{F} = -0.3 \, $, 
$\, \epsilon_{F} = - 0.22 \, $ and 
$\, \epsilon_{F} = -0.14 \, $ ($\delta = 0.5$) 
for the relation (\ref{VolCentrDispRel}) 
(body-centered lattice).}  
\label{Fig17}
\end{figure*}

 In the intervals 
$\, \epsilon_{F} \in \left( \epsilon^{\cal A}_{1} , 
\, \epsilon^{\cal A}_{2} \right) \, $ the Fermi surfaces 
have genus 6 (and rank 3) (Fig. \ref{Fig17}).

 Here we use the intervals 
$\, \left[ \widetilde{\epsilon}_{0} (Q , \delta) , \, 
\widetilde{\epsilon}_{0} (P , \delta) \right] \, $ 
(Fig. \ref{Fig8}) to estimate the intervals 
$\, \left[ \epsilon^{\cal B}_{1} (\delta) , \, 
\epsilon^{\cal B}_{2} (\delta) \right] \, $. 
As we have already said, in the general case we have 
the inclusion
$$\left[ \widetilde{\epsilon}_{0} (Q , \delta) , \,
\widetilde{\epsilon}_{0} (P , \delta) \right]
\,\,\, \subseteq \,\,\, 
\left[ \epsilon^{\cal B}_{1} (\delta) , \,
\epsilon^{\cal B}_{2} (\delta) \right] \,\,\, , $$
moreover, for large Zones $\, W_{\alpha} \, $ these intervals
coincide in order of magnitude (and often coincide exactly).

\begin{figure*}[t]
\begin{center}
\includegraphics[width=\linewidth]{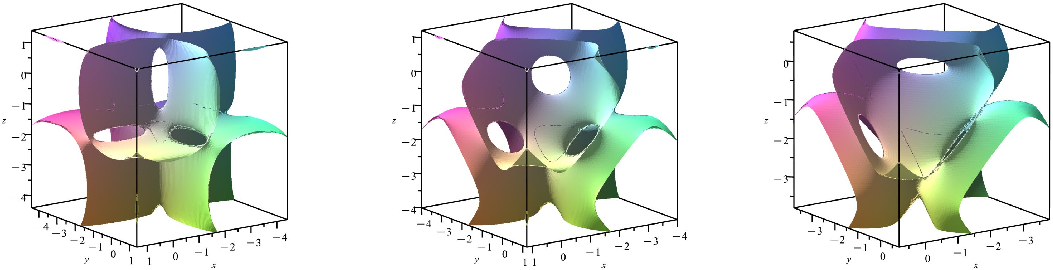}
\end{center}
\caption{The Fermi surfaces
$\, \epsilon_{0.1} ({\bf p}) = - 0.06 \, $, 
$\, \epsilon_{0.5} ({\bf p}) = - 0.2 \, $,
$\, \epsilon_{0.9} ({\bf p}) = - 0.245 \, $ 
for the relation (\ref{VolCentrDispRel})
(body-centered lattice).
}  
\label{Fig18}
\end{figure*}

 Near the intervals 
$\, \left[ \widetilde{\epsilon}_{0} (Q , \delta) , \,
\widetilde{\epsilon}_{0} (P , \delta) \right] \, $, 
the geometry of the Fermi surfaces for the relation 
(\ref{VolCentrDispRel}) can be described relatively simply. 
Namely, the Fermi surfaces here are formed by ``spheroids'' 
located at the nodes of a face-centered lattice 
(note that the lattice inverse to the body-centered lattice 
is face-centered), connected by (very short) cylinders. 
It should be noted that the cross-sections of these 
cylinders have different shapes for different values 
of $\delta$ (Fig. \ref{Fig18}).

\begin{figure*}[t]
\begin{center}
\includegraphics[width=\linewidth]{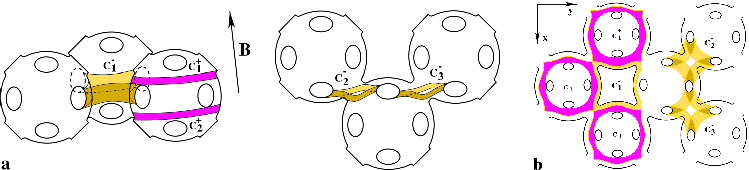}
\end{center}
\caption{(a) Cylinders 
$\, C^{-}_{1}, C^{+}_{1}, C^{+}_{2}, C^{-}_{2}, C^{-}_{3} \, $ 
for a direction $\, {\bf B} \, $, close to $\, \widehat{z} \, $. 
(b) Shape of the cylinders $\, C^{-}_{1}$, $\, C^{-}_{2} \, $ 
and $\, C^{-}_{3} \, $ for $\, {\bf B} \parallel \widehat{z} \, $ 
(top view).}  
\label{Fig19}
\end{figure*}

 For directions of $\, {\bf B} \, $ close to 
$\, \widehat{z} \, $, Fermi surfaces contain 
4 carriers of open trajectories
$\, \left\{ T_{1}, T_{2}, T_{3}, T_{4} \right\} \, $, 
separated by 5 cylinders of closed trajectories 
$\, \left\{ C^{-}_{1}, C^{-}_{2}, C^{-}_{3}, 
C^{+}_{1}, C^{+}_{2} \right\} \, $ 
(Fig. \ref{Fig19}).

\begin{figure}[t]
\begin{center}
\includegraphics[width=0.8\linewidth]{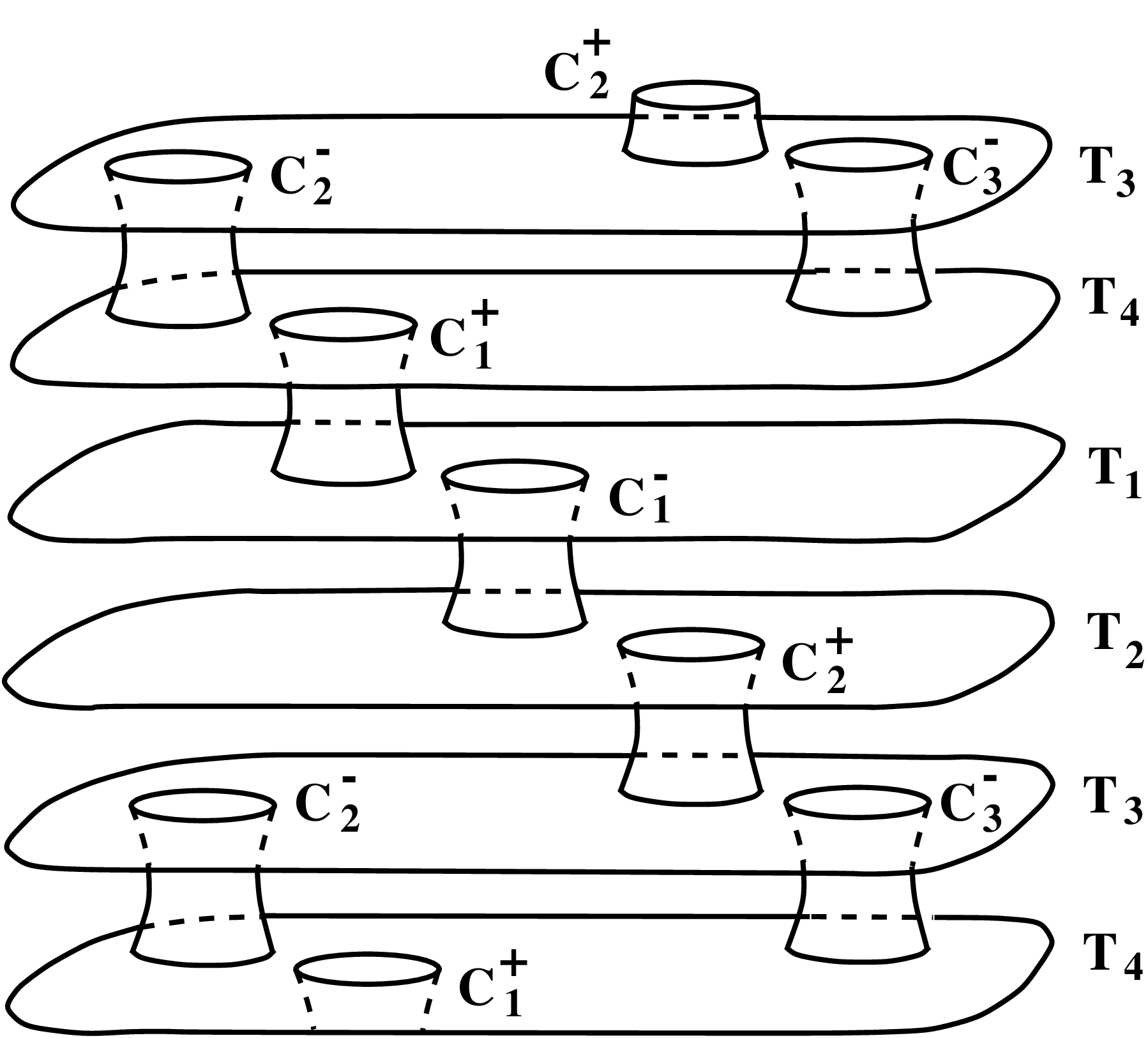}
\end{center}
\caption{Topological diagram of the connection of carriers 
$\, T_{1}, T_{2}, T_{3}, T_{4} \, $ and cylinders 
$\, C^{-}_{1}, C^{-}_{2}, C^{-}_{3}, C^{+}_{1}, C^{+}_{2} \, $ 
into the Fermi surface at $\, {\bf n} \simeq \widehat{z} \, $ 
(body-centered lattice).}  
\label{Fig20}
\end{figure}

 Obviously, the cylinder $\, C^{-}_{1} \, $ is symmetric with 
respect to the reflection $\, {\bf p} \rightarrow - {\bf p} \, $, 
while the cylinders $\, C^{-}_{2} $, $\, C^{-}_{3} \, $, as well as 
$\, C^{+}_{1} $, $\, C^{+}_{2} \, $, transform into each other 
under this reflection. The topological diagram of the connection 
of the carriers $\, T_{1}, T_{2}, T_{3}, T_{4} \, $ and the cylinders 
$\, C^{-}_{1}, C^{-}_{2}, C^{-}_{3}, C^{+}_{1}, C^{+}_{2} \, $ 
is shown in Fig. \ref{Fig20}.

 The disappearance of any of the cylinders upon deviation 
of $\, {\bf n} = {\bf B}/B \, $ from $\, \widehat{z} \, $ 
leads to the destruction of the carriers of open trajectories 
connected by it (the possibility of trajectories jumping 
from one carrier to another appears). This destruction, 
however, can be eliminated by changing the value of 
$\, \epsilon_{F} \, $, if it is not accompanied at the same time
by the disappearance of another cylinder of closed trajectories 
of the opposite type. As we have already mentioned, the 
boundaries of the Zones $\, W_{\alpha} \, $ are therefore 
determined by the disappearance of at least one cylinder of 
the electron type and one of the hole type (and the 
destruction of all carriers of open trajectories).

 Due to symmetry, both hole-type cylinders 
($C^{+}_{1}$ and  $C^{+}_{2}$) disappear simultaneously 
(for the same ${\bf B}$ directions). As a consequence, 
any point on the boundary $\, W_{1} \, $ corresponds to 
the simultaneous disappearance of the cylinders $C^{+}_{1}$ 
and $C^{+}_{2}$.

 In our case, for the impossibility of restoring any of the 
pairs of carriers of open trajectories 
($\left\{ T_{1}, T_{2} \right\}$ or $\left\{ T_{3}, T_{4} \right\}$) 
under a change of the value of $\, \epsilon_{F} \, $, 
the absence of all the cylinders 
$\, C^{-}_{1}$, $\, C^{-}_{2}$, $\, C^{-}_{3}$ is required 
(Fig. \ref{Fig20}).

 Thus, the boundary of $\, W_{1} \, $ is determined by the 
disappearance of the pair 
$ \, \left\{ C^{+}_{1}, C^{+}_{2} \right\} \, $, and also 
either the cylinder $\, C^{-}_{1} \, $, or the pair 
$ \, \left\{ C^{-}_{2}, C^{-}_{3} \right\} \, $ 
(depending on which disappears last).

 From the shape of the Fermi surfaces for different 
$\, \delta \, $ (Fig. \ref{Fig18}), one can see that the 
height of the cylinder $\, C^{-}_{1} \, $ is quite large, 
and the heights of the cylinders $\, C^{-}_{2} \, $ and 
$\, C^{-}_{3} \, $ are quite small, for small values of 
$\, \delta \, $. For $\, \delta \rightarrow 1 \, $ the 
situation is reversed, i.e. the height of $\, C^{-}_{1} \, $ 
becomes small, and the heights of $\, C^{-}_{2} \, $ and 
$\, C^{-}_{3} \, $ increase noticeably. It can be stated, 
therefore, that for small values of $\, \delta \, $ the 
position of points $P$ and $Q$ is determined by the 
disappearance of the cylinders 
$ \, \left\{ C^{+}_{1}, C^{+}_{2}, C^{-}_{1} \right\} \, $, 
and for $\, \delta \rightarrow 1 \, $ - by the disappearance 
of the cylinders 
$ \, \left\{ C^{+}_{1}, C^{+}_{2}, C^{-}_{2}, C^{-}_{3} \right\} \, $.

\vspace{1mm}

\begin{figure*}[t]
\begin{center}
\includegraphics[width=\linewidth]{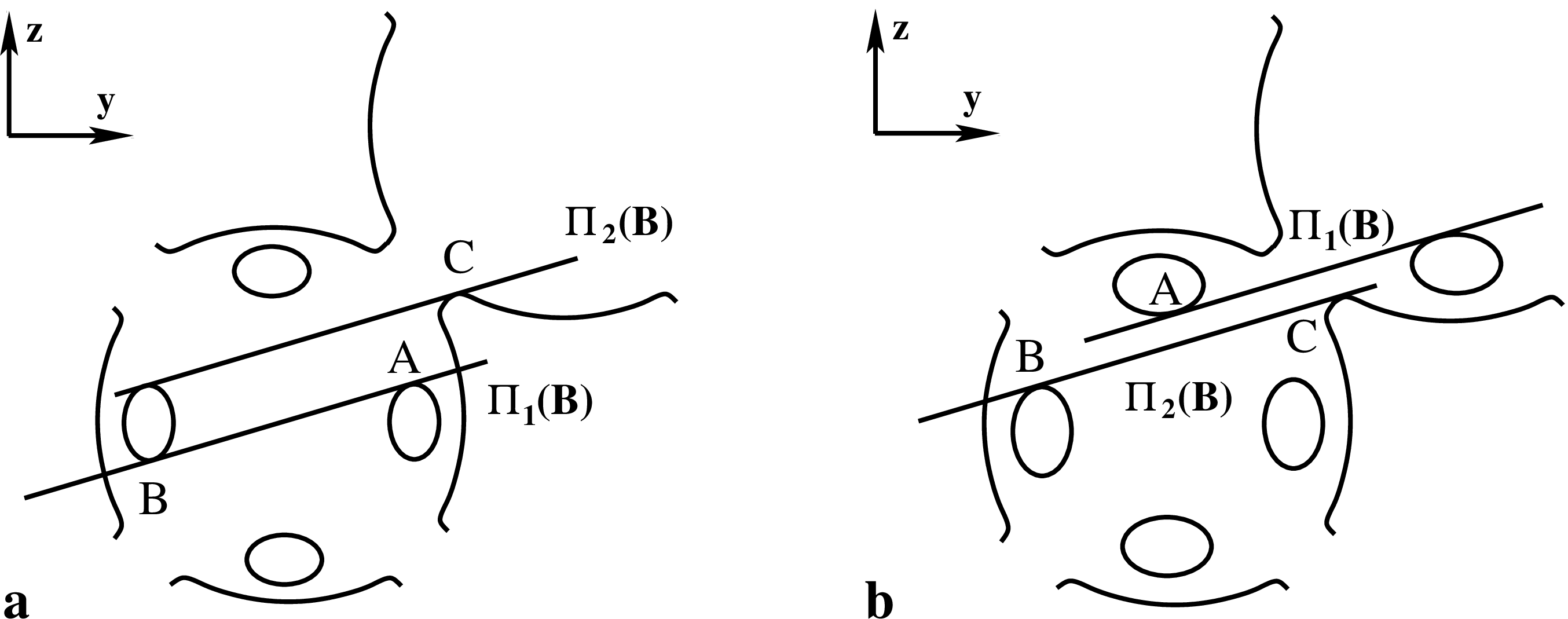}
\end{center}
\caption{(a) Section of the Fermi surface 
$\, \epsilon_{\delta} ({\bf p}) = \widetilde{\epsilon}_{0} (P) \, $ 
by planes $\, \Pi_{1,2} ({\bf B}) \, $ containing cylinders 
$\, C^{-}_{1} \, $ and $\, C^{+}_{1}$, $\, C^{+}_{2} \, $ 
of zero height.
(b) Section of the Fermi surface 
$\, \epsilon_{\delta} ({\bf p}) = \widetilde{\epsilon}_{0} (P) \, $ 
by planes $\, \Pi_{1,2} ({\bf B}) \, $ containing cylinders 
$\, C^{-}_{2} $, $\, C^{-}_{3} \, $ and 
$\, C^{+}_{1}$, $\, C^{+}_{2} \, $ of zero height 
(projection onto the $yz$-plane).
}  
\label{Fig21}
\end{figure*}

 Fig. \ref{Fig21}, a, shows the section of the Fermi surfaces 
$\, \epsilon_{\delta} ({\bf p}) = \widetilde{\epsilon}_{0} (P) \, $ 
by special planes $\, \Pi_{1,2} ({\bf B}) \perp {\bf B} \, $ 
for the point $P$ determined by the disappearance of the cylinders 
$\, C^{-}_{1} $, $\, C^{+}_{1} \, $ and $\, C^{+}_{2} \, $.

\vspace{1mm}

 The plane $\, \Pi_{1} ({\bf B}) \, $ is defined by the 
equation $\, z = \mu y \, $ (with some coefficient $\mu$) 
and is tangent to $\, S_{F} \, $, in particular, at 
points $A$ and $B$ with some coordinates 
$\, \left( b , \, a , \, \mu a \right) \, $ and 
$\, \left( b , \, - a , \, - \mu a \right) \, $.

\vspace{1mm}

 The plane $\, \Pi_{2} ({\bf B}) \, $ is given by the equation
$$\left( z - 2 \mu a \right) \,\,\, = \,\,\, \mu y $$
and touches $\, S_{F} \, $ at a point $C$ with coordinates 
$\, \left( 0 , \, c , \, \mu c + 2 \mu a \right) \, $.

\vspace{1mm}

 We have, therefore,
$$\epsilon_{\delta} \left( b , \, a , \, \mu a \right) 
\,\,\, = \,\,\, \widetilde{\epsilon}_{0} (P, \delta) $$
$$\epsilon_{\delta} \left( 0 , \, c , \, \mu c + 2 \mu a 
\right) \,\,\, = \,\,\, \widetilde{\epsilon}_{0} (P, \delta) $$
$$\nabla \epsilon_{\delta} \left( b , \, a , \, \mu a \right) 
\,\,\, \parallel \,\,\, \left( 0, \, - \mu , \, 1 \right) $$
$$\nabla \epsilon_{\delta} \left( 0 , \, c , \, \mu c + 2 \mu a 
\right) \,\,\, \parallel \,\,\, \left( 0, \, - \mu , \, 1 \right) $$
(touching at the point $B$ occurs automatically), or

\vspace{1mm}
 
\begin{multline*}
\cos b \, \cos a \, \cos \mu a \,\, +  \\
+ \,\, {\delta \over 4} \Big(\cos 2 b \, + \, 
\cos 2 a \, + \, \cos 2 \mu a \Big)  \,\,\, = \,\,\, 
\widetilde{\epsilon}_{0} (P, \delta) \,\,\, ,  
\end{multline*}
\begin{multline*}
\cos c \, \cos (\mu c + 2 \mu a) \,\, +   \\
+ \,\, {\delta \over 4} \Big(1 \, + \, \cos 2 c \, + \,
\cos ( 2 \mu c + 4 \mu a ) \Big)
\,\,\, = \,\,\, \widetilde{\epsilon}_{0} (P, \delta) \,\,\, ,
\end{multline*}
$$\cos a \, \cos \mu a \,\, + \,\, \delta \, \cos b 
\,\,\, = \,\,\, 0 \,\,\, ,  $$
\begin{multline*}
\cos b \, \sin a \, \cos \mu a \,\, + \,\,
\delta \, \sin a \, \cos a  \,\, +  \\
+ \,\, \mu \, \cos b \, \cos a \, \sin \mu a 
\,\, + \,\, \mu \delta \, \sin \mu a \, \cos \mu a 
\,\,\, = \,\,\, 0 \,\,\, ,  
\end{multline*}
\begin{multline}
\label{Psystem1}
\!\!\!\!\!  \mu \, \cos c \, \sin (\mu c + 2 \mu a)
\,\, + \,\, \mu \delta \, \sin (\mu c + 2 \mu a) \, 
\cos (\mu c + 2 \mu a) \,\, +  \\
+ \,\, \sin c \, \cos (\mu c + 2 \mu a) \,\, + \,\,
\delta \, \sin c \, \cos c  \,\,\, = \,\,\, 0   
\end{multline}

\vspace{1mm}

 Fig. \ref{Fig21}, b, shows the section of the Fermi surfaces 
$\, \epsilon_{\delta} ({\bf p}) = \widetilde{\epsilon}_{0} (P) \, $ 
by special planes $\, \Pi_{1,2} ({\bf B}) \perp {\bf B} \, $ for the 
point $P$ determined by the disappearance of the cylinders 
$\, C^{-}_{2} $, $\, C^{-}_{3} \, $, $\, C^{+}_{1} \, $ 
and $\, C^{+}_{2} \, $.

\vspace{1mm}
 
 The plane $\, \Pi_{1} ({\bf B}) \, $ is now given by the equation
$$ z - {\pi \over 2} \,\,\, = \,\,\,  \mu \left( 
y - {\pi \over 2} \right) $$ 
(with some coefficient $\mu$) and is tangent to $\, S_{F} \, $ 
at a point $A$ with some coordinates 
$\, \left( b , \, a , \, \mu a + \pi (1 - \mu)/2 \right) \, $.

\vspace{1mm}

 The plane $\, \Pi_{2} ({\bf B}) \, $ is given by the equation
$$ z  \,\,\, = \,\,\,  \mu  y  \, + \, \lambda $$ 
(with some $\lambda$) and touches $\, S_{F} \, $ at point $C$ 
with coordinates 
$\, \left( 0 , \, c , \, \mu c + \lambda \right) \, $ and 
point $B$ with coordinates 
$\, \left( h , \, d , \, \mu d + \lambda \right) \, $.

\vspace{1mm}

 We have now
$$\epsilon_{\delta} \left( b , \, a , \, 
\mu a  +  \pi (1 - \mu)/2 \right) 
\,\,\, = \,\,\, \widetilde{\epsilon}_{0} (P, \delta) $$
$$\epsilon_{\delta} \left( 0 , \, c , \, \mu c + \lambda
\right) \,\,\, = \,\,\, \widetilde{\epsilon}_{0} (P, \delta) $$
$$\epsilon_{\delta} \left( h , \, d , \, \mu d + \lambda
\right) \,\,\, = \,\,\, \widetilde{\epsilon}_{0} (P, \delta) $$
$$\nabla \epsilon_{\delta} \left( b , \, a , \, 
\mu a   +  \pi (1 - \mu)/2  \right) 
\,\,\, \parallel \,\,\, \left( 0, \, - \mu , \, 1 \right) $$
$$\nabla \epsilon_{\delta} \left( 0 , \, c , \, \mu c + \lambda 
\right) \,\,\, \parallel \,\,\, \left( 0, \, - \mu , \, 1 \right) $$
$$\nabla \epsilon_{\delta} \left( h , \, d , \, \mu d + \lambda 
\right) \,\,\, \parallel \,\,\, \left( 0, \, - \mu , \, 1 \right) $$
or
\begin{multline*}
{\delta \over 4} \Big(\cos 2 b \, + \, 
\cos 2 a \, + \, \cos 2 \left(
\mu a +  \pi (1 - \mu)/2 \right) \Big) \,\, +  \\
+ \,\, \cos b \, \cos a \, \cos \left( 
\mu a +  \pi (1 - \mu)/2 \right) \,\,\, = \,\,\, 
\widetilde{\epsilon}_{0} (P, \delta) \,\,\, ,  
\end{multline*}
\begin{multline*}
{\delta \over 4} \Big(1 \, + \, 
\cos 2 c \, + \, \cos 2 \left(
\mu c +  \lambda \right) \Big) \,\, +  \\
+ \,\,  \cos c \, \cos \left( 
\mu c +  \lambda \right) \,\,\, = \,\,\, 
\widetilde{\epsilon}_{0} (P, \delta) \,\,\, ,  
\end{multline*}
\begin{multline*}
{\delta \over 4} \Big(\cos 2 h \, + \, 
\cos 2 d \, + \, \cos 2 \left(
\mu d  +  \lambda \right) \Big) \,\, +  \\
+ \,\, \cos h \, \cos d \, \cos \left( 
\mu d  +  \lambda \right) \,\,\, = \,\,\, 
\widetilde{\epsilon}_{0} (P, \delta) \,\,\, ,  
\end{multline*}
$$\cos a \, \cos \left( \mu a +  \pi (1 - \mu)/2 \right) 
\,\, + \,\, \delta \, \cos b 
\,\,\, = \,\,\, 0 \,\,\, ,  $$
\begin{multline*}
\cos b \, \sin a \, \cos  
\left( \mu a +  \pi (1 - \mu)/2 \right) \,\, + \,\,
\delta \, \sin a \, \cos a  \,\, +  \\
+ \,\, \mu \Big( \cos b \, \cos a \, 
\,\, + \,\, \delta \, \cos 
\left( \mu a +  \pi (1 - \mu)/2 \right)  \Big)  \times  \\
\times  \,\, \sin \left( \mu a +  \pi (1 - \mu)/2 \right) 
\,\,\, = \,\,\, 0 \,\,\, ,  
\end{multline*}
\begin{multline*}
\mu \, \cos c \, \sin (\mu c + \lambda )
\,\, + \,\, \mu \delta \, \sin (\mu c + \lambda ) \, 
\cos (\mu c + \lambda ) \,\, +  \\
+ \,\, \sin c \, \cos (\mu c + \lambda ) \,\, + \,\,
\delta \, \sin c \, \cos c  \,\,\, = \,\,\, 0  \,\,\, , 
\end{multline*}
$$\cos d \, \cos \left( \mu d +  \lambda \right) 
\,\, + \,\, \delta \, \cos h
\,\,\, = \,\,\, 0 \,\,\, ,  $$
\begin{multline}
\label{Psystem2}
\cos h \, \sin d \, \cos  
\left( \mu d +  \lambda \right) \,\, + \,\,
\delta \, \sin d \, \cos d  \,\, +  \\
+ \,\, \mu \Big( \cos h \, \cos d \, 
\,\, + \,\, \delta \, \cos 
\left( \mu d  +  \lambda \right)  \Big) 
\sin \left( \mu d + \lambda \right) 
\,\, = \,\, 0 
\end{multline}

 By independently solving systems (\ref{Psystem1}) and 
(\ref{Psystem2}) and choosing for each $\, \delta \, $ the 
solution corresponding to the larger value of $\, \mu \, $, 
we obtain the dependence $\, \widetilde{\epsilon}_{0} (P) \, $ 
that we need.

\vspace{2mm}

 For the direction $\, {\bf n} = Q \, $ it is convenient 
to make the substitution
$$x \,\,\, = \,\,\, {v - u \over \sqrt{2}} \,\,\, , \quad
y \,\,\, = \,\,\, {v + u \over \sqrt{2}} $$

 As is easy to verify, in the new coordinates the dispersion 
relation (\ref{VolCentrDispRel}) has the form
\begin{multline*}
\epsilon_{\delta} ({\bf p}) \,\,\, = \,\,\, 
{1 \over 2}  \Big( \cos \sqrt{2} u \,\, + \,\, 
\cos \sqrt{2} v \Big) \cos z   \,\,\, +    \\
+ \,\,\, {\delta \over 2} \, \cos \sqrt{2} u \, \cdot \,
\cos \sqrt{2} v \,\,\, + \,\,\,  {\delta \over 4} \, \cos 2z
\end{multline*}

\vspace{1mm}

\begin{figure*}[t]
\begin{center}
\includegraphics[width=\linewidth]{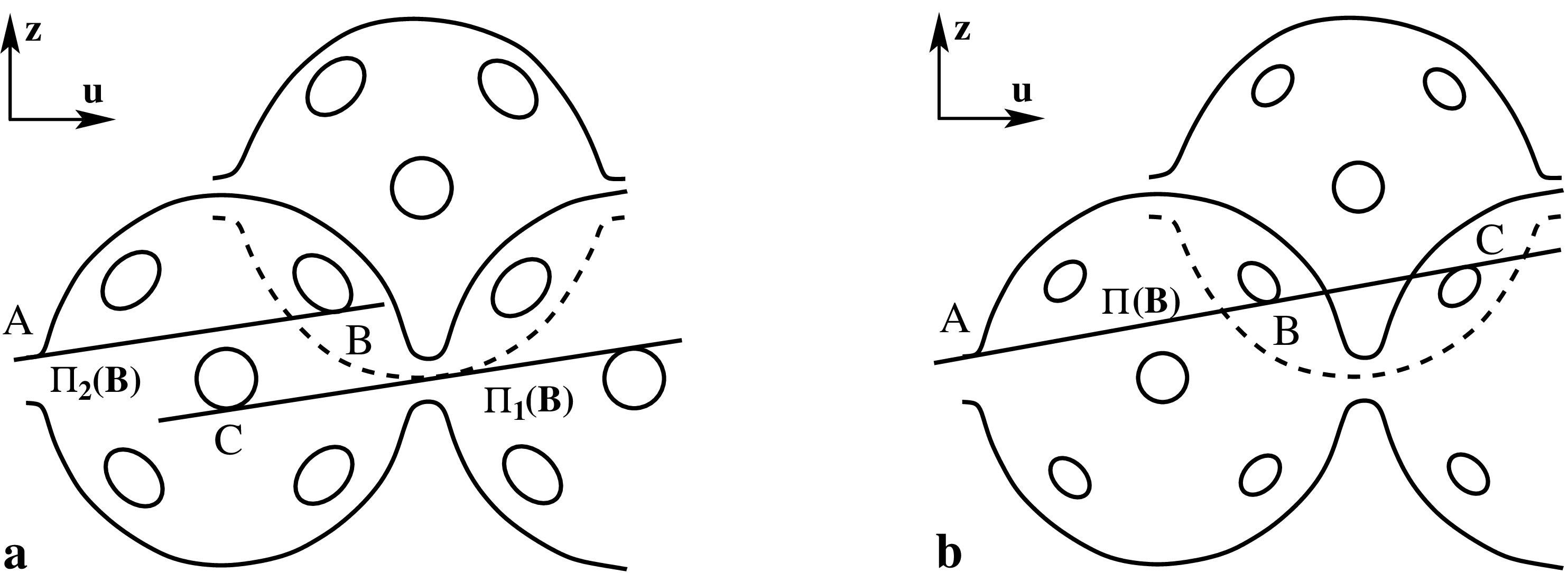}
\end{center}
\caption{(a) Section of the Fermi surface 
$\, \epsilon_{\delta} ({\bf p}) = \widetilde{\epsilon}_{0} (Q) \, $ 
by the planes $\, \Pi_{1,2} ({\bf B}) \, $ containing the cylinders 
$\, C^{-}_{1} \, $ and $\, C^{+}_{1}$, $\, C^{+}_{2} \, $ 
of zero height.
(b) Section of the Fermi surface 
$\, \epsilon_{\delta} ({\bf p}) = \widetilde{\epsilon}_{0} (Q) \, $ 
by the plane $\, \Pi ({\bf B}) \, $ containing the cylinders 
$\, C^{-}_{2}$, $\, C^{-}_{3} \, $ and $\, C^{+}_{1}$, $\, C^{+}_{2} \, $ 
of zero height (projection onto the plane $\, x + y = 0 $).
}  
\label{Fig22}
\end{figure*}

 Fig. \ref{Fig22}, a, shows the section of the Fermi surfaces 
$\, \epsilon_{\delta} ({\bf p}) = \widetilde{\epsilon}_{0} (Q) \, $ 
by special planes $\, \Pi_{1,2} ({\bf B}) \perp {\bf B} \, $ for the 
point $Q$ determined by the disappearance of the cylinders 
$\, C^{-}_{1} $, $\, C^{+}_{1} \, $ and $\, C^{+}_{2} \, $.

\vspace{1mm}

 The plane $\, \Pi_{1} ({\bf B}) \, $ is given in 
new coordinates by the equation
$$z \,\,\, = \,\,\, \nu 
\left( u - {\pi \over \sqrt{2}} \right) $$
(with some $\nu$) and touches $\, S_{F} \, $ at 
point $C$ with coordinates 
$\, \left( c, \, \pi / \sqrt{2} , \, 
\nu (c - \pi / \sqrt{2}) \right) \, $.

\vspace{1mm}

 The plane $\, \Pi_{2} ({\bf B}) \, $ is given by the equation
$$z \,\,\, = \,\,\, \nu  u \, + \, \zeta $$
(with some $\zeta$) and touches $\, S_{F} \, $ at point $A$
with coordinates
$\, \left( a, \, 0 , \, \nu a + \zeta \right) \, $ and at 
point $B$ with coordinates 
$\, \left( b, \, h , \, \nu b + \zeta \right) \, $.

\vspace{1mm}

 The complete system for the problem parameters has the form
\begin{multline*}
{1 \over 2}  \Big( \cos \sqrt{2} c \,\, - \,\, 1 
\Big) \cos \nu (c - \pi / \sqrt{2})  \,\,\, -    \\
- \,\,\, {\delta \over 2} \, \cos \sqrt{2} c \,\,\, + \,\,\,  
{\delta \over 4} \, \cos 2 \nu (c - \pi / \sqrt{2})
\,\,\, = \,\,\, \widetilde{\epsilon}_{0} (Q, \delta)  \,\,\, ,
\end{multline*}
\begin{multline*}
{1 \over 2}  \Big( \cos \sqrt{2} a \,\, + \,\, 1 
\Big) \cos (\nu a + \zeta)  \,\,\, +    \\
+ \,\,\, {\delta \over 2} \, \cos \sqrt{2} a 
\,\,\, + \,\,\,  {\delta \over 4} \, \cos 2 (\nu a + \zeta)
\,\,\, = \,\,\, \widetilde{\epsilon}_{0} (Q, \delta)  \,\,\, ,
\end{multline*}
\begin{multline*}
{1 \over 2}  \Big( \cos \sqrt{2} b \,\, + \,\, 
\cos \sqrt{2} h \Big) \cos (\nu b + \zeta)   \,\,\, +    \\
+ \,\, {\delta \over 2} \, \cos \sqrt{2} b \, \cdot \,
\cos \sqrt{2} h \,\,\, + \,\,\,  {\delta \over 4} \, 
\cos 2 (\nu b + \zeta) \,\, = \,\, 
\widetilde{\epsilon}_{0} (Q, \delta)  \,\, ,
\end{multline*}
\begin{multline*}
\!\!\! \nu \left( \cos \sqrt{2} c - 1 \right) \sin 
\nu (c - \pi / \sqrt{2} ) \, + \, \nu \delta \,
\sin 2 \nu (c - \pi / \sqrt{2}) \,\, +  \\
+ \,\,\, \sqrt{2} \, \sin \sqrt{2} c \, \cos 
\nu (c - \pi / \sqrt{2}) \,\, - \,\, \sqrt{2} \delta \,
\sin \sqrt{2} c \,\,\, = \,\,\, 0 \,\,\, , 
\end{multline*}
\begin{multline*}
\!\!\! \nu \left( \cos \sqrt{2} a + 1 \right) \sin 
(\nu a + \zeta ) \, + \, \nu \delta \,
\sin 2 (\nu a + \zeta) \,\, +  \\
+ \,\,\, \sqrt{2} \, \sin \sqrt{2} a \, \cos 
(\nu a + \zeta) \,\, + \,\, \sqrt{2} \delta \,
\sin \sqrt{2} a \,\,\, = \,\,\, 0 \,\,\, , 
\end{multline*}
$$\cos (\nu b + \zeta) \,\, + \,\, \delta \, 
\cos \sqrt{2} b  \,\,\, = \,\,\, 0  \,\,\, ,  $$
\begin{multline}
\label{Qsystem1}
\!\!\! \nu \left( \cos \sqrt{2} b + \cos \sqrt{2} h
\right) \sin (\nu b + \zeta ) \, + \, \nu \delta \,
\sin 2 (\nu b + \zeta) \,\, +  \\
+ \,\,\, \sqrt{2} \, \sin \sqrt{2} b \, \cos 
(\nu b + \zeta) \,\, + \,\, \sqrt{2} \delta \,
\sin \sqrt{2} b \, \cos \sqrt{2} h \,\,\, = \,\,\, 0 
\end{multline}

\vspace{2mm}

  Fig. \ref{Fig22}, b, shows the section of the Fermi surfaces 
$\, \epsilon_{\delta} ({\bf p}) = \widetilde{\epsilon}_{0} (Q) \, $ 
by a special plane $\, \Pi ({\bf B}) \perp {\bf B} \, $ for the 
point $Q$ determined by the disappearance of the cylinders 
$\, C^{-}_{2} $, $\, C^{-}_{3} \, $, $\, C^{+}_{1} \, $ 
and $\, C^{+}_{2} \, $.

\vspace{1mm}

 The plane $\, \Pi ({\bf B}) \, $ is given by the equation 
$\, z = \nu u + \lambda \, $ (with some $\nu$ and $\lambda$) 
and is tangent to the surface $\, S_{F} \, $ at 5 (non-equivalent) 
points. By virtue of symmetry, to determine the dependence 
$\, \widetilde{\epsilon}_{0} (Q) \, $ we only need to consider 
its tangency at 3 points $A$, $B$ and $C$ with coordinates 
$\, (a_{1}, 0, a_{3}) \, $, $\, (b_{1}, b_{2}, b_{3}) \, $, 
$\, (c_{1}, c_{2}, c_{3}) \, $ 
(their projections are shown in Fig. \ref{Fig22}).

\vspace{1mm}

 The complete system for the problem parameters now has the form
\begin{multline*}
{1 \over 2}  \Big( \cos \sqrt{2} a_{1} \,\, + \,\, 1 \Big) 
\cos a_{3}   \,\,\, +    \\
+ \,\,\, {\delta \over 2} \, \cos \sqrt{2} a_{1} \,\,\, + \,\,\,  
{\delta \over 4} \, \cos 2 a_{3} \,\,\, = \,\,\, 
\widetilde{\epsilon}_{0} (Q, \delta)  \,\,\, ,
\end{multline*}
\begin{multline*}
{1 \over 2}  \Big( \cos \sqrt{2} b_{1} \,\, + \,\, 
 \cos \sqrt{2} b_{2} \Big) \cos b_{3}   \,\,\, +    \\
+ \,\,\, {\delta \over 2} \, \cos \sqrt{2} b_{1} \, \cdot \,
\cos \sqrt{2} b_{2} \,\,\, + \,\,\,  
{\delta \over 4} \, \cos 2 b_{3} \,\,\, = \,\,\, 
\widetilde{\epsilon}_{0} (Q, \delta)  \,\,\, ,
\end{multline*}
\begin{multline*}
{1 \over 2}  \Big( \cos \sqrt{2} c_{1} \,\, + \,\, 
 \cos \sqrt{2} c_{2} \Big) \cos c_{3}   \,\,\, +    \\
+ \,\,\, {\delta \over 2} \, \cos \sqrt{2} c_{1} \, \cdot \,
\cos \sqrt{2} c_{2} \,\,\, + \,\,\,  
{\delta \over 4} \, \cos 2 c_{3} \,\,\, = \,\,\, 
\widetilde{\epsilon}_{0} (Q, \delta)  \,\,\, ,
\end{multline*}
$$\nu \,\,\, = \,\,\, {b_{3} - a_{3} \over b_{1} - a_{1}} 
\,\,\, = \,\,\, {c_{3} - b_{3} \over c_{1} - b_{1}} \,\,\, , $$
$$\cos b_{3} \, + \, \delta \cos \sqrt{2} b_{1} 
\,\,\, = \,\,\, \cos c_{3} \, + \, \delta \cos \sqrt{2} c_{1} 
\,\,\, = \,\,\, 0  \,\,\, , $$
\begin{multline*}
\sqrt{2} \sin \sqrt{2} a_{1} \, \cdot \, \cos a_{3} \,\, + \,\, 
\sqrt{2} \delta \sin \sqrt{2} a_{1} \,\, +   \\
+ \,\, \nu \sin a_{3} \Big( \cos \sqrt{2} a_{1} \,\, + \,\, 1 
\Big) \,\, + \,\, \nu \delta \sin 2 a_{3} \,\,\, = \,\,\, 0
\,\,\, , 
\end{multline*}
\begin{multline*}
\sqrt{2} \sin \sqrt{2} b_{1} \, \cdot \, \cos b_{3} \,\, + \,\, 
\sqrt{2} \delta \sin \sqrt{2} b_{1} \, \cdot \, \cos \sqrt{2} b_{2} 
\,\, +   \\
+ \, \nu \sin b_{3} \Big( \cos \sqrt{2} b_{1} \, + \, 
\cos \sqrt{2} b_{2} \Big) \, + \, \nu \delta \sin 2 b_{3} 
\,\, = \,\, 0  \,\, , 
\end{multline*}
\begin{multline}
\label{Qsystem2}
\sqrt{2} \sin \sqrt{2} c_{1} \, \cdot \, \cos c_{3} \,\, + \,\, 
\sqrt{2} \delta \sin \sqrt{2} c_{1} \, \cdot \, \cos \sqrt{2} c_{2} 
\,\, +   \\
+ \, \nu \sin c_{3} \Big( \cos \sqrt{2} c_{1} \, + \, 
\cos \sqrt{2} c_{2} \Big) \, + \, \nu \delta \sin 2 c_{3} 
\,\, = \,\, 0  
\end{multline}

 By independently solving systems (\ref{Qsystem1}) 
and (\ref{Qsystem2}) and choosing for each $\, \delta \, $ 
the solution corresponding to the larger value 
of $\, \nu \, $, we obtain the dependence
$\, \widetilde{\epsilon}_{0} (Q) \, $ that we need.

\vspace{1mm}
 
\begin{figure*}[t]
\begin{center}
\includegraphics[width=0.9\linewidth,height=0.6\linewidth]{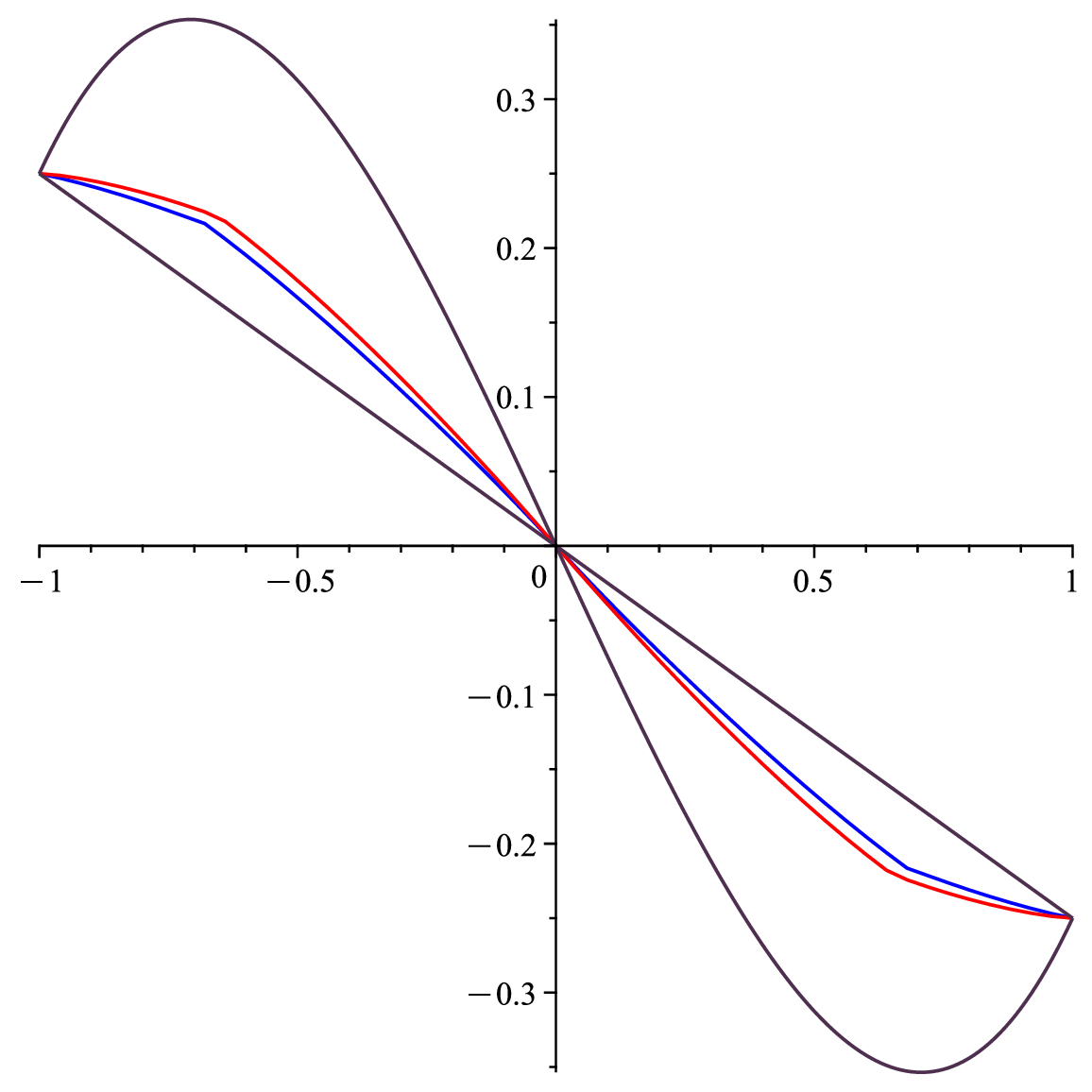}
\end{center}
\caption{Boundaries of the intervals
$\, \left[ \widetilde{\epsilon}_{0} (Q , \delta) , \,
\widetilde{\epsilon}_{0} (P , \delta) \right] \, $,
$\, \left[ \widetilde{\epsilon}_{0} (P , \delta) , \,
\widetilde{\epsilon}_{0} (Q , \delta) \right] \, $ and
$\, \left( \epsilon^{\cal A}_{1} (\delta) , \, 
\epsilon^{\cal A}_{2} (\delta) \right) \, $ 
for the relation (\ref{VolCentrDispRel}) in the range 
$\, - 1 < \delta < 1 \, $ (body-centered lattice).}  
\label{Fig23}
\end{figure*}

 Fig. \ref{Fig23} shows the values of the functions
$\, \widetilde{\epsilon}_{0} (P) \, $ and 
$\, \widetilde{\epsilon}_{0} (Q) \, $, as well as the 
boundaries of the interval 
$\, \left( \epsilon^{\cal A}_{1} (\delta) , \, 
\epsilon^{\cal A}_{2} (\delta) \right) \, $ in the range 
$\, - 1 < \delta < 1 \, $. It can be seen that the width 
of the interval 
$\, \left[ \widetilde{\epsilon}_{0} (Q , \delta) , 
\, \widetilde{\epsilon}_{0} (P , \delta) \right] \, $ 
does not exceed here $0.1$ of the width of the interval 
$\, \left( \epsilon^{\cal A}_{1} (\delta) , 
\, \epsilon^{\cal A}_{2} (\delta) \right) \, $. 
It can also be seen that the maximum width of the 
interval 
$\, \left[ \widetilde{\epsilon}_{0} (Q , \delta) , \, 
\widetilde{\epsilon}_{0} (P , \delta) \right] \, $ 
(for $\, \delta \simeq 0.6$) does not exceed $1 \%$
of the width of the interval 
$\, \left[ \epsilon_{\min} (\delta) , \, 
\epsilon_{\max} (\delta) \right] \, $ for the 
dispersion law (\ref{VolCentrDispRel}).

\section{Conclusion}
\setcounter{equation}{0}

 We estimate the width of the energy interval corresponding 
to the emergence of ultra-complex conductivity diagrams 
for analytical dispersion relations arising in conductors 
of cubic symmetry in the tight-binding approximation. The 
study uses higher corrections to the leading approximation 
in the tight-binding limit, which allows us to estimate 
the widths of this interval for the cases of simple and 
body-centered cubic lattices. The width of this interval 
is directly related to the probability of the occurrence 
of ultra-complex conductivity diagrams in materials described 
by the tight-binding approximation. In addition, the study of 
the dependence of the specified interval on the spectrum 
parameters also allows us to evaluate the possibility of 
obtaining ultra-complex conductivity diagrams and observing 
the effects associated with them by means of external 
influence on the sample under study.

\vspace{5mm}

This work was supported by the Russian Science Foundation 
under grant no. 26-11-00292, 
https://rscf.ru/project/26-11-00292/

\end{document}